# Mitigating the quantum hype

Olivier Ezratty [1]

[1] consultant and author of the Understanding quantum technologies ebook, Paris, France, olivier@oezratty.net @olivez

We are in the midst of quantum hype with some excessive claims of quantum computing potential, many vendors' and even some research organizations' exaggerations, and a funding frenzy for very low technology readiness level startups. Governments are contributing to this hype with their large quantum initiatives and their technology sovereignty aspirations. Technology hypes are not bad per se since they create emulation, drive innovations and also contribute to attracting new talents. It works as scientists and vendors deliver progress and innovation on a continuous basis after a so-called peak of expectations. It fails with exaggerated overpromises and underdeliveries that last too long. It could cut short research and innovation funding, creating some sort of quantum winter. After looking at the shape and form of technology and science hypes and driving some lessons from past hypes, we investigate the current quantum hype and its specifics. We find that, although there is some significant uncertainty on the potential to create real scalable quantum computers, the scientific and vendor fields are relatively sane and solid compared to other technology hypes. The vendors hype has some profound and disruptive impact on the organization of fundamental research. Also, quantum technologies comprise other fields like quantum telecommunications and quantum sensing with a higher technology readiness level, which are less prone to hype. We then make some proposals to mitigate the potential negative effects of the current quantum hype including recommendations on scientific communication to strengthen the trust in quantum science, vendor behavior improvements, benchmarking methodologies, public education and putting in place a responsible research and innovation approach.

## INTRODUCTION

Artificial intelligence specialists who have been through its last "winter" in the late 1980s and early 1990s keep saying that quantum computing, if not quantum technologies on a broader scale, are bound for the same fate: a drastic cut in public research spendings and innovation funding. Their assumption is based on observing quantum technology vendors and even researchers overhype, on a series of oversold and unkept promises in quantum computing and on the perceived slow improvement pace of the domain.

The quantum race launched by many governments, particularly with the USA, China, and other developed countries in between is also artificially fueling this trend, fed by "technology sovereignty" concerns. The recent large funding rounds of leading startups like IonQ, PsiQuantum and Rigetti contributed to entertain this overhype perception. Some go as far - way too far - with arguing that quantum computing is a scam created by scientists who found a way to get funding for their research ventures.

Trying to stage a balanced view, this paper describes the shape and form of this quantum hype and what are its similarities and differences with other digital era hypes like symbolic artificial intelligence, 3D television, consumer 3D printing, virtual and augmented reality, Blockchain and crypto-currencies as well as with other science related hypes. It proposes some insights and code of conduct for the quantum ecosystem that would avoid the pitfalls of the current quantum hype while keeping the benefits of a vibrant scientific and technology ecosystem. It also builds on and complements earlier work by science philosophers in the field of quantum ethics and responsible innovation[1].

Quantum fundamental research has been active since 1900 with Planck's fundamental discovery of the energy quanta. Since then, the only sluggish period was World War II when American and German scientists like Enrico Fermi and Werner Heisenberg were mobilized on nuclear physics[2]. It enabled the creation of the first nuclear bombs and their unfortunate usage. It was one the several cases where innovation was not synonymous with progress, creating a philosophical chasm between science and society. This "crisis of science" still resonates with scientists.

The industry payback of quantum physics research came with the transistor invention in 1947, the laser in the early 1960s, and many other technology feats (TVs, LEDs, GPS, ...) leading to the "first quantum revolution" based digital era we are enjoying today.

The "second quantum revolution" era deals with controlling individual quantum objects (atoms, electrons, photons) and using superposition and entanglement. The early 1980s were defining moments with Yuri Manin and Richard Feynman expressing their ideas to create respectively gate-based quantum computers and quantum simulators (in 1980 and 1981) and then, with Alain Aspect *et al* undertaking their famous 1982 Bell inequalities violation experiment using distant entangled photons[3].

We are now 40 years ahead, and even though quantum science advances have been continuous, usable quantum computers offering a quantum advantage compared with classical computers are not there yet and this can be the source of some impatience[4]. However, three other applications of this second quantum revolution are alive and well: quantum telecommunications, cryptography and sensing. To some extent, the latter is well under-hyped.

# HYPE ONTOLOGY

Hype is a term referring to over mediatization and inflated, excessive or misleading claims that are applicable to particular products, products categories, technology trends, scientific domains, personalities like artists or politicians, and even speculative financial bubbles and scams[5]. Hype characterizations range from broad societal phenomena to explicit, well thought-out and planned marketing strategies. It can be confused with marketing exaggerations which are among its implementation artefacts. On the other hand, buzz is a quiet way to disseminate promotional content, before it gets amplified by hype.

Hype existed way before the Internet but it can now be orchestrated by leveraging social media to amplify their effects[6]. As a marketing strategy, hype can be applied in various fields like in fashion to promote new styles and brands. It can even be based on creating artificial scarcity or fake strong demand. Also, many old Ponzi financial schemes were created with hype development mechanisms.

## Science and technology hypes

Scientific and technology hypes do not involve the same stakeholders as fashion, financial and politics hypes. They deal a lot with how the scientific community, the industry and society interact with each other in an unorganized fashion. As we get closer to the commercial world, business and financial values and systems are becoming powerful hype echo chambers.

Technology hypes are not bad per se[7]. It depends on their scale and how self-fulfilling promises are delivered. Hypes drive research, invention and innovation on a global basis[8]. Their main use case is to attract government and private sector funding. They can indirectly help make progress with science and make the field attractive to new talent. And unlike Gene Kranz's famous quote, failure *is* an option. No failure would mean that not enough scientific and technology avenues were investigated. The field of quantum technologies is probably exemplifying this phenomenon with a sheer diversity of pursued technology options, and not just with the many quantum computing qubit types that are investigated.

Scientific hype can also happen way before it enters the entrepreneurial and commercial scene, when various positive or beneficial aspects of science are inappropriately exaggerated and sensationalized, with the caveat that evaluation appropriateness relies on value judgements[9]. Hype can show up first in scientific papers' titles created by research labs communicators[10]. This frequently happens in life science with the past examples of pluripotent stem cells and cancer curing monoclonal antibodies[11]. This can be driven by the way researchers are funded and rewarded in most countries. Technology vendors are indeed not the only ones competing with each other for funding. Science is also a very competitive field, where visibility, recognition, careers as well as public and private funding are at stake.

The hype can then be amplified when science communication and academic publishing is translated in layman's terms in news media[12]. It can also be a side effect of papers being published on pre-print servers like Arxiv, without being peer reviewed although most scientific papers benefit from some media visibility after they are published in peer-reviewed publications.

## Hype emotions and irrationality

Science and technology hype is a field of collision between, on one hand, information streams stoking emotions and irrationality, and on the other hand scientific, technologic and even business rationality. It builds on strong beliefs in science driven progress and on the confusion between laboratory experiments and production-grade solutions. Hype goes way beyond a more classical build-up of rational expectations and a societal contract on the required investments needed to deliver value[13].

Emotionally, hype drives hope, envy and fear. Hope of solving key problems like with healthcare or climate change. Envy and fear of missing out on becoming wealthy (FOMO) for entrepreneurs and investors, or, for governments, of being overpowered by another country. Lastly, fear of losing competitiveness or missing business opportunities for corporations.

Hype related emotions are also easier to manipulate given the ignorance by its various target audiences of the various scientific or technology obstacles in creating actual solutions. It can build on magical thinking and occasionally use science-fiction references[14].

## Gartner hype curve model

Technology hypes were practically defined by the Gartner Group with its famous hype curve model, created in 1995[15]. It tries to capture new technologies' visibility and success cycles in some predictive way. The model uses a non-linear curve with unscaled time in X and visibility or expectations in Y. After a new technology appearance trigger, the first peak of visibility corresponds to a hype-driven "peak of inflated expectations" when some highly positive buzz is amplified by news media, frequently at the border line of magical thinking. This buzz can be created and fed by a variable mix of scientists, analysts, consultants, influencers, entrepreneurs, corporations and sometimes, governments themselves. This is where we are right now with quantum computing as shown in Figure 1.

Then, if and when over-expectations are not matched by actual technology capacities and benefits, trust vanishes in a "trough of disillusionment" with negative news coverage and an overall lack of confidence in the technology and its creators. This gap between expectations and the actual delivery capabilities of science and technology can drive disinvestments in the related science or technology, similarly to what happened with artificial intelligence during its two winters in the early 1970s and 1990s.

It can also happen when some technology innovation is not creating real perceived societal progress and value.

At last, when technology finally matures, its visibility shines back with a growing "slope of enlightenment" and a "plateau of productivity". The technology has then the potential to become mainstream and commoditized in the marketplace, whether in the enterprise or the consumer space depending on the technology.

This Gartner hype curve is an over-simplistic and empirical model that does not capture well what can happen with disruptive science and technologies[16]. It is constructed with adding an initial "buzz" Bell curve and a technology maturity S curve, given these do not rely on the same metric. It is missing scientific, technology and even business rationales. It deals mainly with the emotional response to new science and technology and is void of any rationality. It also presupposes that a topic's visibility, expectations and needs are correlated. Its premise is that at some point in time, the given technology domain will succeed. But what if science and technology fail to deliver? What if the new needs the innovation is supposed to address do not really exist? The curve is also built out of a strong survival bias, forgetting the various technologies that totally failed for one reason or the other[17].

The market is now flooded with a growing number of technology trends and fads. How is this affecting visibility? Sometimes, the technology trend is not well defined, like "nanotechnologies", which drives the market crazy[18]. The length of the trough of disillusionment can be very long, spanning several decades. We are still in it for symbolic AI and it may be what will happen with scalable quantum computing[19].

One strong shortcoming of the hype curve is that Gartner is using it to advise corporations on when to adopt new technologies. In a rationale way, it should not be done because it is popular or not done because it is not in the media radar[20]. Innovation winners are frequently contrarians!

The only rationale behind becoming a follower is when technology adoption is requiring a strong ecosystem of applications and services. But it could be quantified without resorting to empirical noise estimations.

New technologies and scientific domains can thus traverse other cycles as presented in Figure 1. Some new technologies succeed without relying on inflated hype. That was the case with micro-computers, which had no winter between their beginnings in the mid-1970s and their broad consumer adoption in the late 1990s. There is also a high uncertainty grey zone in between two opposite situations: hype fulfillment works over the long term through some slow multi-decades progress and the unenviable unrecovered failures graveyard. This can help build some scenarios. Quantum computing could enter and stay in this grey zone for the next two decades (scenario 5), unless significant scientific and technology breakthroughs are demonstrated in the next five years (scenario 4).

Other innovation models can be considered but are less used to assess the effects of hype and do not really take into account deep tech innovations' scientific and technology uncertainty. Geoffrey Moore's innovation model with its chasm and tornado describes the way innovations become mainstream thanks to their adoption by consecutive user segments regardless of technology readiness, as shown in Figure 2[21]. It acutely describes the chasm happening between technology early adopters and the early majority that follows-up (or not) and ensures that the offer later becomes mainstream. This corresponds to the barrier to technology commoditization. When it is overcome, the market grows quickly with ecosystems' expansion and distribution fluidification, sometimes with significant price decrease. This period is the theater of a "tornado" during which dominant positions are consolidated, up to creating monopolies when solutions rely on large ecosystems.

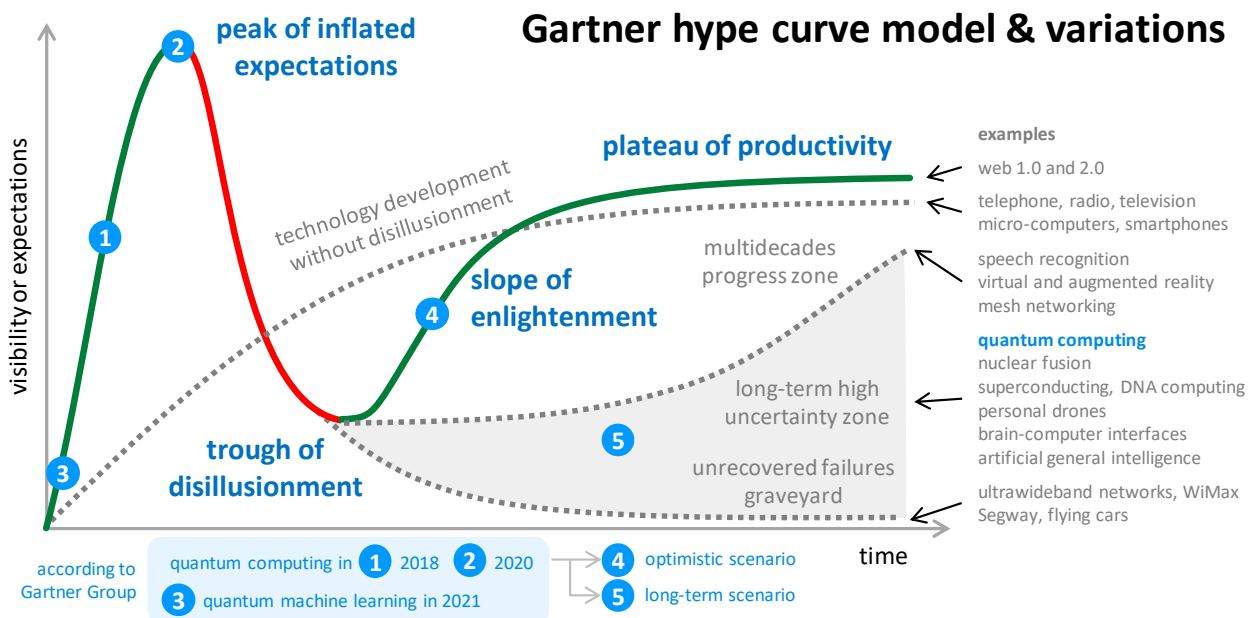

Figure 1 : Gartner hype curve model and some alternate routes.

# Geoffrey Moore innovation adoption model

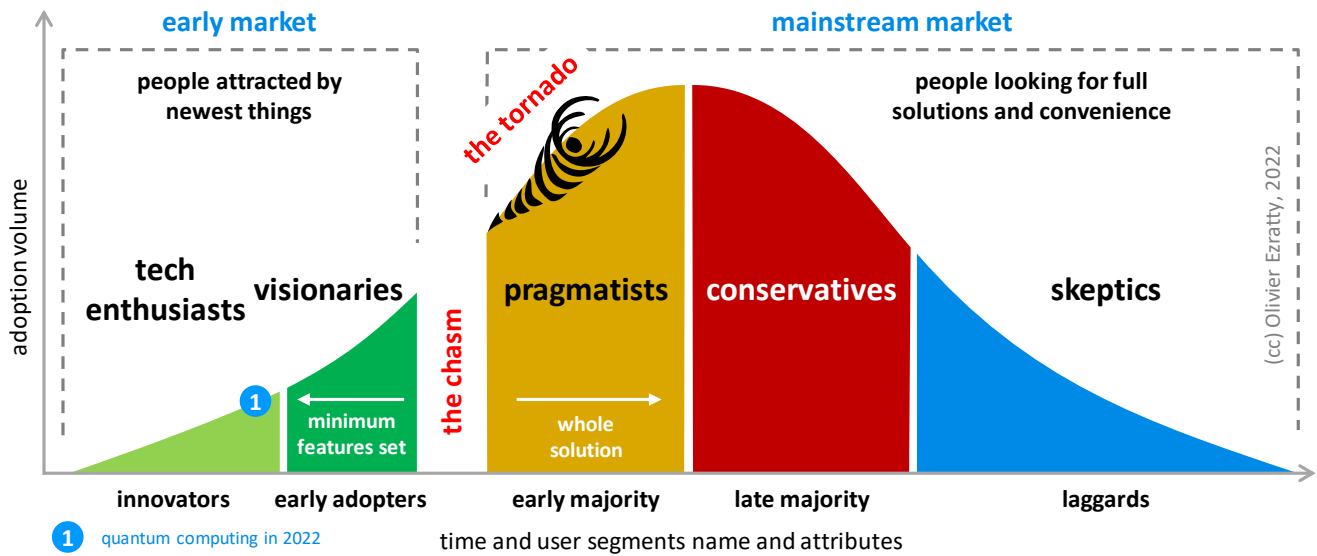

**Figure 2 : Geoffrey Moore's innovation adoption model.**

If we were to use Moore's model with quantum computing, we'd rather say that we are still at the innovators stage with existing pre-NISQ quantum processing systems that are still experimental devices[22]. The minimum feature set corresponds to the ability to reach some quantum advantage to solve practical business problems. Early adopters will jump on the bandwagon with useful NISQ or large scale and fault-tolerant computing systems in an undetermined timeframe and then, when the software and cloud ecosystem grow, we'll cross the chasm and see the early majority adopt quantum computing massively.

Clayton Christensen's innovation disruption model covers innovations that are appealing to low-end or unserved consumers and then become mainstream and concurrently are based on innovative business models[23]. The model works when a new technology and its implementations significantly expand a given market. The classical examples are the personal computer and the smartphone. It is less applicable to quantum computing given it will, at least at the beginning, be a sub-market of the narrow high-performance computing market.

In both cases, Clayton Christensen's and Geoffrey Moore's models are describing the implementation details of the slope of enlightenment in the Gartner hype model, when technology starts to actually work. In the case of quantum computing, these considerations will become useful when quantum computers really scale and provide some quantum advantage compared to classical computers. But some aspects of Moore's innovation adoption model are put in place early on, such as software tools ecosystems and platform like IBM Qiskit. It prepares the market early on for a dissemination of software solutions and skills when scalable quantum hardware actually starts to work and can become a leading platform.

## HISTORY LESSONS AND ANALOGIES

The last decades saw an explosion of digital and other technology waves, most of them successfully deployed at large scale. Micro-computers invaded the geek world, then the workspace and at last, our homes. Besides the Millennium Bug overhyped fears, Y2K marked the beginning of the consumer digital era, starting with web-based digital music, digital photography, digital video and television, e-commerce, the mobile Internet, the sharing economy, all sorts of disintermediation services and at last cloud computing.

There were however some failures, with hype waves and adversarial outcomes. In many cases, while these hypes peaked, some skepticism could be built out of common sense. In Figure 3, I provide a rough comparison of successful and failed technology hypes with some simple explanation of their relative outcomes.

I will look here at some product categories and not about particular products that succeeded and crashed or crashed right away (Betamax, Altavista, MySpace, Blackberry, Segway, Theranos). There are also many case studies I won't investigate like hydrogen fuel, electric vehicles and precision medicine.

### Symbolic artificial intelligence

In the 1980s, symbolic AI and expert systems were trendy but had practical implementation issues. Capturing expert knowledge was difficult and could not be automatized when contents were not massively digitized as they are today. There were even dedicated machines tailored for running the artificial intelligence LISP programming language.

It could not compete with generic CPU based machines from Intel and the likes that were benefiting from a sustained (Gordon) Moore's law. As a result, the couple related startups failed even before they could start selling their hardware. Then, expert systems went out of fashion.

AI rebirth came with deep learning, an extension of machine learning created by Geoff Hinton in 2006. The comeback can be traced to 2012 when deep learning could benefit from improved algorithms and powerful GPUs from Nvidia. The current AI era is mostly a connectionist one, based on machine and deep learning probabilistic models, as opposed to symbolic AI based on rules engines.

As an extension of the old symbolic AI winter, autonomous cars are also progressing quite slowly. The scientific and technology challenges are daunting, particularly when dealing with the interaction between autonomous cars and humans, whether they are walking or in other 2, 3, 4 or more wheels vehicles.

Nowadays, artificial intelligence is an interesting field with regards to how ethics, responsible innovation and environmental issues are considered, mostly as an afterthought. The learning here is, the sooner these matters are investigated and addressed, the better. We will cover that later in the responsible innovation section.

| technology trend | what was oversold | what happened | why it failed or succeeded |
|---|---|---|---|
| symbolic AI | market size, corporate adoption | LISP startups closure, fading trend | hard to capture expert knowledge, systems limitations |
| internet 1.0 | market adoption, business models | technology financial bubble burst | lack of PCs and broadband telecom, decline of interest rates |
| 3D printing | market size, consumer adoption | remains a professional business | no real consumer use case |
| Internet of things | market size | became quietly mainstream | low cost, business use cases |
| 3D television | market size, consumer adoption | nearly abandonned | physiological limits, content production costs, devices costs, market fragmentation |
| VR/AR and metavers | market size, commoditization | became a niche market | |
| smartphones | it would replace desktops | broad usage success, didn't kill desktops | telecom infrastructure, generic platforms and market place |
| machine learning | it would solve all problems and destroy jobs | broad market adoption, no jobs destroyed | generic use cases with data, vision, NLP; dedicated hardware |
| cryptocurrencies | it would take over the world and replace fiducial currencies | did not succeed broadly yet, but its usage seems to grow steadily | confusion between exchange currency and speculative asset, consumer value, complicated |
| blockchains | market adoption | NFT in consumer space, slow enterprise projects otherwise | not an interoperability standard => poor network effect, complicated |
| hyperloops | fast deployments, fast city to city commuting | still at prototype stage | lots of technical issues, infrastructure costs |
| passenger drones | mass and fast deployments | still at prototype stage | energetic cost, infrastructure, air traffic control and safety concerns |
| autonomous cars | when and where it would work | it doesn't work yet in complex environments | AI limits, sensors data fusion, handling interactions with humans |

Figure 3 : table comparing (summarily) various technology hypes characteristics.

### Consumer 3D printing

In the early 2010s, some analysts predicted that home 3D printing would thrive. It did not. Back then, you could simply ask yourself: how many single material plastic pieces do you need to print each and every year in your home that you can't easily buy on Amazon or elsewhere that would justify buying a $300 device and the long printing hours and associated tries and failures? So, failure in the consumer market came from a lack of practical value and, to some extent, from the inexistence of multi-materials 3D printing systems (plastic, steel, electronics, glass).

The 3D printing market is still growing steadily, but mostly in professional markets and in the manufacturing industry (automotive, aerospace, ...). We are far from "The Zero Marginal Cost Society" envisioned by Jeremy Rifkin in 2015. 3D printing has not been consumerized. This is a good learning for those who forecast some consumer adoption of quantum computing.

### 3D TV and VR/AR

Some relatively recent failures in the consumer electronics space could be attributed to some physiological issues. 3D TV was not reproducing a real stereoscopic view experience. It is the same with the current clunky virtual and augmented reality (VR/AR) headsets.

Google tried it with its Google Glass in 2013. It was less clunky, but at the price of a very low resolution of 640x360 pixels and a small angle view of a mere 13°. As a consumer product, Google Glass was sold only between 2013 and 2015 (see bottom of Figure 4).

Back then, I found out how virtual reality and augmented reality were constrained, not really by the available computing power but more by the law of classical optics. Finding a way to send a high-resolution wide-angle image into your eyes is a very complicated classical optical challenge. Nobody has found a solution, even Magic Leap or Microsoft with its Hololens 2 (top of Figure 4)! Yet, Magic Leap raised a total $2.6B, more than half of all quantum startups altogether. It recently repositioned itself in the healthcare market, abandoning the broader consumer market.

It is no brainer to be relaxed when some are advocating the creation of Chief Metaverse Officers in corporations to run their presence in virtual worlds[24]. During the Linden Labs "Second Life" frenzy peaking in 2013, some corporation human resources departments created virtual spaces to hire young professionals, like if they were pre-selecting avatars! It is also not difficult to guess why 3D and VR/AR face some hurdles with contents production costs. The recent rebranding of these virtual worlds into metaverses will not change much the picture.

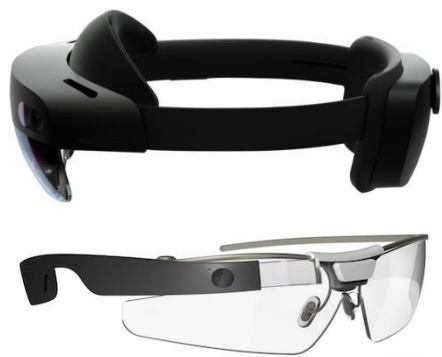

<span style="color:#4A77C0">**Figure 4 : augmented reality headsets from Microsoft (Hololens 2) and Google (Glass Enterprise Edition).**</span>

### Blockchain and cryptocurrencies

We do not know yet if it will succeed or fail. But in any case, it probably will not be due to sophisticated scientific reasons. It could be due to a lack of economies of scale, interoperability issues, regulatory initiatives like China's recent ban of cryptocurrencies and mining[25] or to a lack of societal acceptance or, simply, real consumer value.

The Blockchain is not an interoperability standard per se since you have as many Blockchains as use cases. It prevents the creation of a global platform ecosystem like the web or even smartphone application stores, which Google and Apple consolidate. The cryptocurrencies and Blockchain startups hype is currently much broader than the quantum hype.

In Q2 2021 alone, Blockchain startups raised $4B, so an equivalent of all quantum startups life-to-date and $17B in the first 6 months of 2021[26]. One reason for these high funding levels is that financial sector startups are very attractive to investors compared to deep tech startups. These have the potential ability to create faster revenue streams and become more valuable.

As a subtrack of the Blockchain market, the Non-Fungible Token (NFT) created its own hype with amazing funding levels reached by Sorare ($739M, France) and Dapper Labs ($607M, Canada). These weird NFT-based virtual objects are used to create unique digital collectibles for fans of sports games, art and the likes. Sorare offers "*a global fantasy football game where players can buy, trade, and play with official digital cards*". Here, the only sciences at stake are social engineering and user experience design! Hacking human emotions is much easier than getting rid of quantum decoherence and scaling qubits.

More recently, the Blockchain and cryptocurrency waves were wrapped into the "web3" concept that envisions a decentralized web to fight the dominance of the Facebook, Google and Amazons of this world[27]. Again, science is not at stake, but more voodoo-economics and some computing architecture that can easily be deconstructed.

Lastly, there's a significant difference compared to the quantum hype: no developed country is (significantly) funding public research in the Blockchain, cryptocurrencies, NFT and web3 spaces. It makes sense given these technologies are designed to deconstruct the control of governments over currencies. Only a few countries like El Salvador or Barbados embraced the Bitcoin as an official currency[28]. Other countries are trying to regulate it.

### Science hypes

The famous Gartner hype curve can move ahead or stop after the trough of disillusionment. When it stops as described in Figure 1, it can be related to some societal and economical reason (like there is no real business case, people do not like it, it is not a priority, it is too expensive) or to scientific showstoppers (it does not work yet, it is unsafe, etc.). In the science realm, where are the overpromises and hypes that led to some sort of technology winter?

You can count nuclear fusion in, and it is still a long-term promise, although being back in fashion with a couple startups launched in that domain[29]. These have raised more than $2B so far[30]. Being able to predict when, where and how nuclear fusion could be controlled to produce electricity is as great a challenge as forecasting when some scalable quantum computer could be able to break 2048 bit RSA keys. In a sense, these are similar classes of challenges.

Unconventional computing brought its own fair share of failed hypes with optical computing and superconducting computing. We have been hearing for a while about spintronics and memristors that are still in the making. These are mostly being researched by large companies like HPE, IBM, Thales and the likes.

DNA-based precision medicine and genome engineering has also been oversold for about 10 years now. Some confusion was made between DNA sequencing advances and its derived applications in genomics-based therapies. It is still a promising field. We then have a small new hype with DNA-based long term data storage.

These cases are closer to the quantum hype in several respects: their scientific dimension, fact-checking toughness, low technology readiness levels and scientific uncertainty. They are also driving a lower awareness in the general public than non-scientific hypes like cryptocurrencies, NFTs and metaverses[31].

## QUANTUM HYPE CHARACTERIZATION

The current quantum computing hype actually started in 1994 when Peter Shor, then at the Bell Labs and now at the MIT, created his famous eponymous integer factoring quantum algorithm that could potentially break most public-key-based encryption systems[32].

That year also marked the advent of the consumer web with Netscape's creation. Their IPO in 1995 launched the first Internet hype and startups funding wave, which collapsed in 2000 together with large technology values. It created a short-lived Internet winter before the "web 2.0" rebirth around 2004, and later, with the beginning of the mobile Internet and smartphone era in 2007 when the first Apple iPhone was launched[33].

Shor's algorithm created some excitement with governments, particularly in the USA. The potential to be the first country to break Internet security codes was appealing, as a revival of the famous Enigma machine during World-War II. The difference is you are replacing German submarines!

As the web and mobile Internet usage exploded, starting in the mid to late 2000s, the interest in quantum technologies kept growing. The Snowden 2013 revelations on the NSA Internet global spying capabilities had a profound impact on the Chinese government. It triggered or amplified their willingness to better protect their sensitive government information networks. It may have contributed to their deployment of a massive quantum key distribution network of 2,000 km between Beijing and Shanghai. This network is currently extended to other large cities with an additional 30,000 km of dedicated fiber and endpoints. They also implemented similar capacities using satellites as early as 2017, although with very low key-rates. One has to remember that these networks are dedicated to sharing secret encryption keys. The encrypted data like the video feeds demonstrated between China and Austria in 2017 is still travelling in a classical manner along regular fibers or any wireless links[34].

In the early 2000s, research labs started to assemble a couple of qubits, starting with 2 to 5 qubits between 2002 and 2015. Google's Sycamore "quantum supremacy", although overstated and overinterpreted, was a landmark event that contributed to amplify the quantum computing hype in 2019[35]. We now have up to 127 superconducting qubits in the USA (IBM), but these are way too noisy to be practically usable together[36]. With different methods and the "boson sampling" technique, Chinese researchers went further with 113 detected photon qubits that are phase parametrizable and thus, potentially programmable[37].

Some describe this trend as exponential but it is not really growing as fast. We are currently stuck with fewer than 30 practically usable noisy qubits in the classical gate-based programming model, at least with superconducting and trapped ions technologies[38].

Many challenges remain to be addressed to create usable and scalable quantum computers, one of which involving assembling a large number of physical qubits to create logical "corrected" qubits. This engenders a large overhead creating significant scalability challenges both at the quantum level and at the classical level, with electronics, cabling, cryogenics and energetics. It is a mix of scientific, technological and engineering challenges, the scientific part being stronger than with most other technologies. Other researchers and vendors are trying to exploit existing noisy systems (NISQ, noisy intermediate scale quantum) with noise-resistant algorithms and quantum error mitigation. There is not yet a shared consensus on the way forward. To some extent, it is good news. We still need some fault-tolerance in the related scientific and innovation process.

Still, the quantum computing frenzy is now unbridled. The perceived cryptography threat from quantum computing and Shor's algorithm is still at its height. See for example the TechCrunch article title from 2018 in Figure 5 and an acute description of the phenomenon in MIT Technology Review in Figure 6.

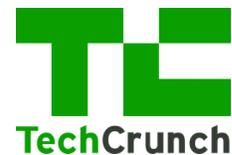

**The quantum computing apocalypse is imminent**

Shlomi Dolev    January 2018

**Figure 5 : example of overstatement of the "imminent" threat created by quantum computing cryptography.**

It drove the creation of post-quantum cryptography systems (PQC) running on classical hardware. NIST is in the final stages of its PQC standardization process which should end by 2024. Many organizations will then be mandated to massively deploy PQCs. We are in a situation equivalent to those folks who built their own home nuclear shelters during the cold war, reminiscent of some advanced form of technology survivalism.

Meanwhile, current security systems are regularly leaking massive personal and sensitive data, mostly due to human originated errors, failures or voluntary leaks. PQC will not address such ongoing weaknesses.

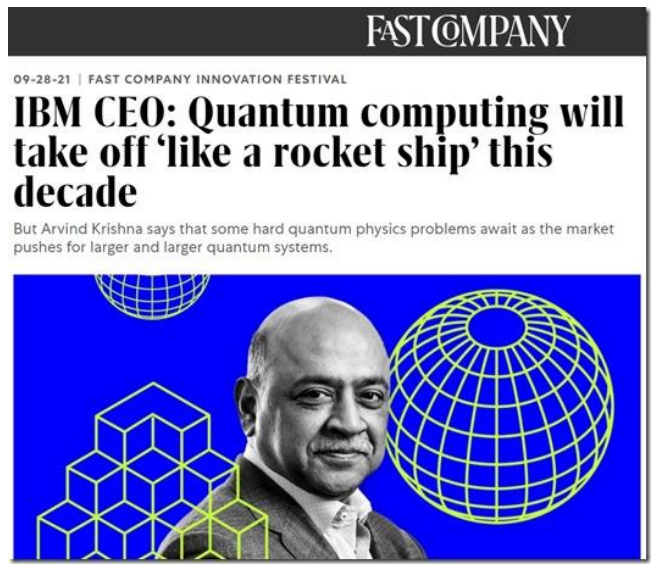

**Quantum Computing Paranoia Creates a New Industry**

Even though quantum computers don't exist yet, security companies are preparing to protect against them.

MIT Technology Review

by Tom Simonite    January 30, 2017

**F**ear sells in the computer security business. And in late 2015 Massachusetts-based Security Innovation got an unexpected boost from one of the scariest organizations around—the National Security Agency.

**Figure 6 : how quantum computing is creating a new paranoia.**

**Most advertised use-cases are overpromises**

One key dimension of the quantum hype shows up with the quantum computing use cases that are usually promoted by vendors and analysts. The classics are ab-initio/de-novo cancer curing drugs and vaccines development, solving earth warming with $CO_2$ capture, creating higher-density, cleaner and more efficient batteries and Haber-Bosch process improvements to create ammonium more energy-efficiently. Of course, all these Earth-saving miracles will be achieved in a snap, compared to billion years of classical computation. Most of these promises are very long term, given the number of reliable (logical) qubits they would require[39]. Some folks even go as far as mentioning weather forecasting as a potential use case, which is really far-fetched[40].

In the real world, scientists and vendors publish weird benchmarks related to solving very exotic physical problems known by only a couple of specialists, like simulating the physics of some rare materials, particularly with ferromagnetism. Large vendors' CEOs also embellish this overpromise land, like IBM's CEO in a 2021 interview for FastCompany, adding a new "managing agricultural risk" use case[41] (see Figure 7). What is that? Maybe thanks to potential fertilizers production yields improvements. This creates self-sustained hype as can be seen in this piece from the Israeli startup Classiq, touting that "quantum computing is the future and it's coming soon"[42]. This was not based on some quantum computing technology assessment but on a simple poll of US IT (information technology) specialists. It was a "meta-poll" (see Figure 8).

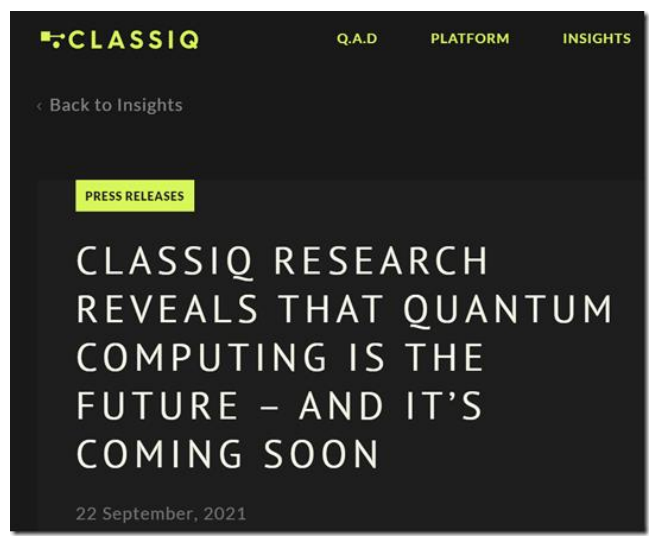

**FAST COMPANY**

09-28-21 | FAST COMPANY INNOVATION FESTIVAL

**IBM CEO: Quantum computing will take off 'like a rocket ship' this decade**

But Arvind Krishna says that some hard quantum physics problems await as the market pushes for larger and larger quantum systems.

**Figure 7 : IBM's CEO fueling the quantum computing hype.**

These messages serve in foot-in-the-door sales tactics to create some sense of urgency with corporate customers[43]. It is maybe for the wrong reasons. Instead of delivering short-term business value like some are abusively promising, it should be about investigating the field of complex problems solving, how it can be implemented now and in the future, and quantum technology readiness for mid-term time horizons.

**CLASSIQ RESEARCH REVEALS THAT QUANTUM COMPUTING IS THE FUTURE – AND IT'S COMING SOON**

22 September, 2021

**Figure 8 : how a ClassiQ survey becomes a truth.**

Overpromises are also frequently conveyed in the growing list of "quantum business conferences" organized and sponsored by vendors as well, as by a new breed of business consultants who position themselves as quantum computing "business cases developers" for large corporations. Right there, right now! It is ironic to see them talk about Corporate adoption of some technology that doesn't really work yet. They want to reach out to business leaders and bypass people in charge of technology in target corporations, an excellent way to sell snake oil to audiences that cannot fact-check any technical claim. One difference with consultants in other domains is that the probability they have no idea what they are talking about is higher in quantum computing.

**Inconsistent signals abound**

The quantum computing scene is prone to fairly inconsistent signals, particularly for appreciating the real state of the art. On one hand, research labs and vendors tout so-called quantum supremacy and advantages, showcasing quantum computers that are supposedly faster than the best supercomputers in the world. On the other hand, we also hear that "real" quantum computers do not exist yet and could appear only in about 10 to 20 years. Reconciling these contradictory messages and fact-checking these claims is far from being easy. You have to look into the details of these quantum supremacies and advantages to understand that they do not relate to any real practical computing advantage. The most famous ones do not use any input data and are more or less random physical quantum systems that are always hard to digitally simulate in a classical computer, a bit like the $1.67 \times 10^{21}$ $H_2O$ molecules in a single drop of water. The table in Figure 9 summarizes this[44].

The same can be said of qubit number growth charts which confuse characterized qubits (like the 10ish qubits from IonQ and Honeywell or 27 qubits from IBM) and prototypes that were never openly characterized, like Google's Bristlecone 72 qubits in 2018 or Rigetti's 128 qubits announced the same year[45]. Some tech publications are also confusing the number of qubits and the Quantum

Volume for a given quantum processor. A Quantum Volume is not a number of qubits[46]! Checking vendor roadmaps' credibility is a tough task as well, such as IBM's hardware roadmap published in September 2020, that planned for 127 qubits in 2021, 433 in 2022 and 1121 in 2023. IBM's Eagle 127 qubit system was introduced in November 2021, right on time. Its basic characterization data like qubit lifetimes and two-qubit gates and readout fidelities are available. Although these are not stellar, it is more or less aligned with the previous 27 and 65 qubits generation systems. For superconducting qubits, it is in itself a prowess. IBM even quietly disclosed the system Quantum Volume, a mere 64, meaning only 6 qubits are useful for decent programming, a regression compared with the Quantum Volume of 128 from their 27 qubits system, corresponding to 7 useful qubits. IBM's roadmap however contradicts another older one, stating a doubling of their qubits' quantum volume every year[47]. Knowing that this corresponds to adding a single operational qubit each and every year, you wonder about how it relates to more than doubling the total number of qubits every other year. To their credit, they recently did a little better than doubling their quantum volume every year[48].

Many scientists also publish significant progress related to quantum error corrections but their practical impacts like the required number of physical qubits to create logical qubits are not necessarily explicit. It also lacks evaluations of the computing time overhead. The same can be said of new hybrid and NISQ algorithms where a quantum advantage is not easy to assess, and when it does exists, can be achieved only on a case by case basis.

| who and when | architecture | algorithm | input data | comment |
|---|---|---|---|---|
| **Google**, Oct 2019 | Sycamore, 53 superconducting qubits | cross entropy benchmarking | **none** | running a random gates algorithm |
| **China**, December 2020 | 70 photons modes GBS (Gaussian Boson Sampling) | interferometer photons mixing | **none** | running a random physical process |
| **IBM Research**, December 2020 | IBM 27 superconducting qubits | symmetric Boolean functions | SLSB3 function parameters | **theoretical demonstration of quantum advantage** |
| **Kerenidis, Diamanti** et al, March 2021 | multi-mode photon dense encoding of verified solution | Quentin Merlin Arthur based verification | output from some quantum computation (not implemented) | **no actual computing done in the experiment** |
| **China**, April 2021 | Quantum walk on 62 superconducting qubits | simple quantum walk | simulating a 2-photons Mach-Zehnder interferometer | **no quantum advantage at all** |
| **University of Arizona**, May 2021 | supervised learning assisted by an entangled sensor network | variational algorithm, classical computing | data extracted from three entangled squeezed light photonic sensors | **not a quantum « computing » advantage per se** |
| **China**, June 2021 | 66 superconducting qubits and 110 couplers, Zuchongzhi 1, then 2.1 | cross entropy benchmarking | **none** | 56 used qubits |
| **China**, September 2021 | | | | 60 used qubits |
| **China**, June 2021 | 144 photons modes GBS and up to 113 detected events | interferometer photons mixing | **none** | **parametrizable photon phases could lead to a programmable system** |

**Figure 9 : table comparing quantum supremacies and advantages claims.**

You must also count with quantum simulators which are programmed in a different way and could well bring some practical advantage before it happens with gate-based quantum computers. This variety of quantum computing paradigms make it quite hard to assess the state of the art and make sound and generic comparisons.

As a result of these mixed and conflicting signals, many non-specialists extrapolate these fuzzy data and make rather inaccurate, particularly overoptimistic, interpretations of quantum computers readiness for prime-time usage, like when a quantum computer will be able to break an RSA-2048 encryption key and threaten Internet's whole security.

**Wild market predictions are made**

Digital technology gold rushes are usually magnified by overly optimistic market predictions. That was the case with many recent technology waves like the Internet of Things, virtual and augmented reality, consumer 3D printing and machine learning-based artificial intelligence.

Most of these predictions are overestimating mid-term markets size, by up to one order of magnitude. You will find in Figure 10 an amazing example with a 2016 IDC prediction, sizing the 2020 virtual and augmented reality markets to $162B[49]. In 2019, IDC downsized its prediction for 2020 to $18.8B! The 2020 actual market size ended up being about $16B, at 10% of the initial prediction[50]. We've seen in a previous section why this prediction was so much off the mark.

One can wonder how analysts come up with these wild market forecasts. There is no safe methodology. In the corporate information technology space, it is commonplace to ask CIOs (chief information officers), CDOs (chief digital officers) or whatever they are called now, to estimate when they will start to invest in new technologies and if possible how much. Since they have no idea and do not want to look like laggards, they provide optimistic yet fuzzy answers. Analysts then tend to forget that most information technology (IT) budgets are at best stable and only prone to rather small reallocations, leaving not much room for new technology investments. IT liability and applications backlogs are still consuming a great share of these budgets. This is the harsh reality of Corporate IT. And in the consumer space, market forecasts are even more difficult to consolidate.

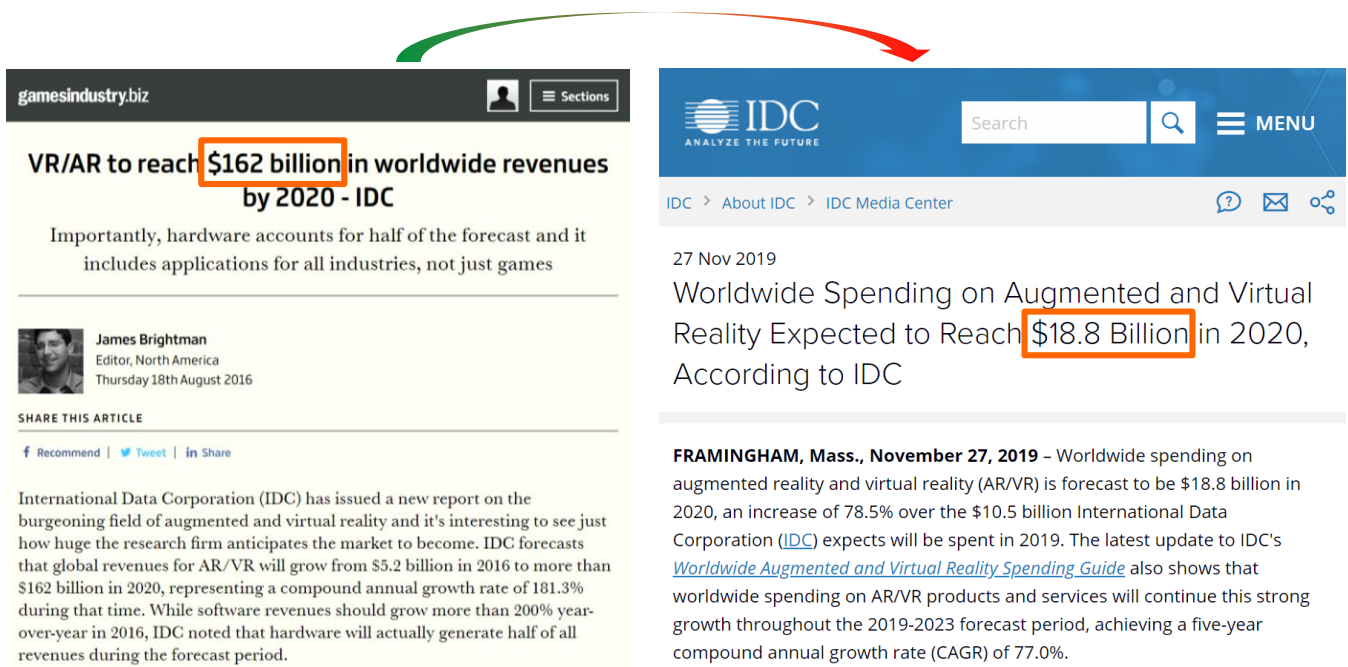

Figure 10 : IDC virtual and augmented reality market prediction and how it was completely off the mark.

Such predictions were also produced for quantum computing market sizes. One difference of these from other tech-related predictions is that they are dependent on a much higher scientific uncertainty. The inconsistent scientific signals discussed previously feed the predictions.

Estimating the size of the quantum computing market looks like reading out the state of a qubit if not in tea leaves: it is highly probabilistic. The quantum computing market is supposed to reach $830M in 2024 for Hyperion Research (in

2021[51]), $1,9B in 2023 for CIR and $2,64B in 2022 for Market Research Future (in 2018). Then we have $8,45B in 2024 for Homeland Security (in 2018), $10B in 2028 for Morgan Stanley (as of 2017), $15B by 2028 for ABI Research (2018) and $64B by 2030 for P&S Intelligence (in 2020).

Research And Markets predicted in May 2021 that the global quantum technology market would even reach $31.57B by 2026, including $14.25B for quantum computing[52]. The discrepancy of these predictions is amazing. Some forecasts can reach other crazy heights. For Bank of America, quantum technologies will be as important as smartphones (see Figure 11). The main reason? Its potential applications in healthcare. One strong caveat: the analysis behind these predictions estimates that quantum computers will handle big data efficiently, which is actually one of their weakness, even in the "scalable quantum computer" realm[53].

The Tell

## Quantum computing will be the smartphone of the 2020s, says Bank of America strategist

Published: Dec. 12, 2019 at 2:40 p.m. ET

By Chris Matthews

Exponentially more computing power may revolutionize health care and cybersecurity

**Figure 11 : Bank of America overstating the impact of quantum computing.**

But sky is a fuzzy limit! As of early 2020, McKinsey predicted that quantum computing would be worth one trillion dollars by 2035 (i.e. $1000B)[54]. This forecast' bias comes from a trick used a few years ago to evaluate the size of Internet of things, artificial intelligence and the hydrogen fuel markets[55]. It is not a market estimation for quantum technologies as such, but the incremental revenue and business value it could generate for businesses, such as in healthcare, finance or transportation verticals. It is a bit like evaluating the software market by summing up the total revenue of the companies who happen to use some software! This would be quite a large number and a significant share of worldwide GDP. The $1T McKinsey market sizing has been exploited again and again in the media and even by financial scammers[56]. One can wonder whether McKinsey anticipated that side effect.

Market predictions should focus on IT products, software and services and be compared with existing reference markets. For example, the 2020 worldwide server market size was $85.7B according to IDC[57]. The real software market could reach $581B in 2021, including $228B in enterprise software, according to Statista[58].

In a 2021 publication, BCG sized the quantum computing market at 20% of its estimated generated value with customers, ending with a $90B to $170B market captured by technology providers, including software and services... some day after 2040, and a more reasonable $1B to $2B before 2030 and $15B to $30B after 2030[59].

So, we have here an uncertainty based on an unknown estimated with some fuzzy technology capability predictions and with a random business value sharing model. In the same vein, The Quantum Daily recently published some Quantum Cloud as a Service market forecast with $26B by 2030. It tried to document its methodology by reminding us that it was based on vendor questionable roadmaps[60].

The best prediction would be to say: "*we have no idea*" (of the market size in 5, 10, 15 or 20 years), but it is obviously harder to sell. It would be more relevant, but costly, to assess the viability of the various vendors scientific and technology roadmaps.

### Vendors are reshuffling the research scene

One other key feature of the current quantum hype comes from a phenomenon that has been commonplace since the first Internet wave from 1995-2001: huge investments from venture capital in a couple visible startups.

Although quantum related venture capital funding does not match the most extreme ones from the classical digital worlds like Uber and its $25.2B venture capital funding, PsiQuantum and IonQ funding each exceeding $600M were good indication of investors' FOMO syndrome (fear of missing out) and their willingness to make significant bets on startups they perceive as being future quantum computing leaders.

Want real hype? Look at Quibi ($1.4B, USA), a failed short online video platform, DoorDash ($2.1B, USA), a mobile food delivery service, Faraday Future ($3.2B, USA/China), an electric vehicles company tediously competing with Tesla and Juul ($15.1B, USA), which sells nicotine vaporizers. A single nicotine vaporizer startup raised more than three times as much as all quantum startups life to date. And the funding was not allocated to fundamental research on nicotine related health risks! These high levels of funding and associated valuations are explainable by the abundance of VC money coming from financial assets reallocations that started a couple years ago.

In the quantum space, this frenzy was exemplified by the appropriation of the technosphere entrepreneurial codes by quantum computing entrepreneurs. IonQ's investor presentation disclosed in March 2021 was an archetype of a set of messages tailored for venture capital funds (VCs), found on a post from Scott Aaronson who is sometimes described as "the conscience" of quantum science"[61]. On top of showcasing the most optimistic market size prediction with a $65B total addressable market in 2030, one claim is that their qubits are "running at room temperature"[62]. A simple search "IonQ cryostat" helps you find that they are using a combination of cryogeny, ultra-high vacuum and laser-based atoms cooling, and thus, do not really run at ambient temperature[63]. The same claim is heard with photon qubits vendors like Orca Computing who avoid mentioning that their photon sources and detectors are cooled at temperatures between 2K and 10K.

This funding trend can be explained by several phenomena: the lack of funding of ambitious consolidated research programs in the public space, the willingness of quantum researchers to create their own venture with more freedom, the abundance of money in the entrepreneurial scene and the quantum hype itself. As a result, quantum startups are competing for PhD-level talent with public research laboratories.

However, the funding frenzy in quantum technology is still very reasonable compared to what happened and still happens in the digital realm. The total quantum technologies startup funding did not exceed $4B worldwide. This is pocket money compared to the $300B invested worldwide in startups just in 2020, which doubled to $621B in 2021[64].

But under our nose, these startups significant funding create a very unique new situation. Due to their low technology readiness level (TRL), quantum computing startups are mostly private fundamental research labs who also happen to undertake some technology developments[65]. In the classical digital world, most startups funding cover product developments, industrialization, ecosystems creation and, above all, customer acquisition and housekeeping costs[66].

The resulting scales are mind-blowing for academic researchers. Startups raising about $20M can create research teams that are bigger than most publicly funded research teams with over 50 PhDs, and with real well paid jobs, not short-term post-docs tenures. The couple startups who raised more than $100M created teams with hundreds of highly skilled researchers and engineers.

You also have to boil in the significant quantum R&D investments from the large US IT vendors which I'll nickname the "IGAMI" (IBM, Google, Amazon, Microsoft, Intel)[67]. IBM has over 400 people working on quantum computing, followed by the others. However, their visibility and wealth can give a false impression of scientific and technology leadership, at least, compared with well-funded startups[68]. This phenomenon is much stronger in the USA than in China, where Alibaba, Tencent and Baidu invest in quantum computing but seemingly more modestly.

In some regions, this phenomenon can create a huge brain drain on skilled quantum scientists. It drives structural changes on how quantum computing research is organized. Academic research organizations now have a harder time attracting and keeping new talent. In many countries where there are not many opened full-time researchers jobs positions, PhDs and post-docs who are looking for a well-paid and full-time job are obviously attracted by the booming private sector.

In the vendors realm, quantum sensing is less talked about and not subject to any visible form of hype. Market predictions are more modest here than with quantum computing. This field is highly fragmented but more mature, with use cases not yet reaching potential mass adoption. It still has some potential for that respect, like in transportation and healthcare. It explains why no quantum sensing startup is making headlines with some record funding. All the largest startup funding went to computing hardware vendors because it is perceived as being the most promising and massive market.

## Many scientists feel the pain

The quantum hype is starting to bother many quantum scientists. They observe, mostly silently, all the above-mentioned claims and events and can feel dispossessed. They know scalable quantum computing is way more difficult to achieve than what vendors are promising. The "Quantum Bullshit Detector" Twitter account between 2019 and 2021 was created to counter outlandish claims coming from all sides[69]. Some scientists even estimate that the way to make it work successfully has not been found. They fear that this overhype could generate some quantum winter like the one that affected artificial intelligence in the early 1990s after the collapse of the expert systems wave. And it would affect other quantum related fields like quantum sensing.

In his "Quantum Computing Hype is Bad for Science" piece published in July 2021, Victor Galitski from the University of Maryland sees "*many "quantum" startups and initiatives, which are promoted and often led by individuals with no relevant expertise or education*" and "*an army of "quantum evangelists," who can't write the Schrödinger equation (or maybe don't even know what it is), but promote the fairy tales about the bounties of quantum computing revolution*"[70]. As a result, "*talent migrates from legitimate science into quantum evangelism with no content*".

When PhDs and post-docs move out of public research labs to create their startup, these are usually serious technology ventures even though they are still doing fundamental research and driven by wishful thinking. They sometimes hire an external CEO with a business background, who plays an "evangelism" role. Most of the time, this is the marketing side of rather serious startups that are still in their R&D and product development phase.

But scientists made entrepreneurs may also feel obliged to leave good scientific methods and practices aside and, for example, favor secret against openness, and as, a result, avoid peer reviewing. This is one of the reasons why it's so difficult to objectively compare the various early quantum computers created by industry vendors, from the large ones to startups with a common set of benchmark practices. It even can lead to announcements of quantum computer deployments with no data at all on qubits number and quality, like what OQC did in the UK in July 2021 with their first cloud deployment[71]. Others go as far as claiming the creation of hundreds of qubit chipsets, without providing any details, these being not connected with each other with multi-qubit gates, nor characterized at all. Victor Galistki even estimates that regular classical computers are being falsely presented as quantum computers sitting in the cloud. This is possible but, I hope and presume, anecdotal.

But all this is not just about marketing and communication. If it is fluffy and still delivers in the end, everything is fine. But otherwise, things could indeed fall back. I got such feedback from some corporate quantum specialists who fear that overpromises and underdeliveries could cut-off their efforts in a couple of years.

The quantum community must put things in order, to avoid getting trapped like this by its own shortcomings. As John Preskill said at the Q2B conference organized in December 2021 by QC-Ware, *"There is a line between setting ambitious goals and fanning inflated expectations... for us as a community, we'll be better off if we try to stay on the right side of that line"*. The only thing is... who defines where the line is? It also requires more quantum scientists to be outspoken in media, including social media, and events to explain science progresses and challenges as well as fight fake science news[72]. Even if they are not necessarily accustomed to spending time in media or have a general background on quantum technologies. But it is possible to learn both. Many of them have this potential[73]. It also circles back to the educational challenge on quantum science and technologies.

### QUANTUM HYPE SPECIFICS

We have just looked at rather commonplace phenomena so far that describe the current quantum hype. How different is the quantum hype phenomenon compared to the various other hypes described in the History lessons and analogies section? There are many differences, mostly related to the scientific dimension and diversity in the second quantum revolution. Also, quantum research works relatively well and is making continuous progress.

#### Quantum uncertainty is mostly scientific

The largest difference is the nature of the uncertainty that is mainly a scientific one, not a matter of market adoption dynamics or some physiological limitations. It is even surprising to see such hype being built so early with regards to the technology maturity, at least for quantum computing. This is an upside-down situation compared to the crypto world: there is market demand but a low technology readiness level (in quantum computing) versus a moderate market demand and a relatively high TRL (in cryptos *aka* Bitcoins and others).

Quantum computing is a complicated domain. You need to delve into quantum physics, linear algebra formalism, microwaves electronics and lasers, thermodynamics, cryogeny, quantum error corrections, quantum algorithms, complexity theories and the likes. Understanding it "full-stack" is quite a challenging task. And engineering will not save us.

The scientific uncertainty of quantum computing creates time-scale collisions. The time-scale of science is long term while that of technology and business is rather short-term. The inconsistency starts to show up with quantum startups adopting short-term practices in a long-term science field or even when industry customers prototype quantum algorithms without the relevant hardware to run it in production scale[74]. On top of that, the source of quantum computing speedups as compared to classical computing is even not clear and agreed upon!

#### Quantum research is reproducible

In 2005, John P. A. Ioannidis, a Stanford University professor published his seminal paper "Why Most Published Research Findings Are False" [75]. He described the many statistical biases he found in research publications about drugs clinical trials, mainly related to innovative cancer treatments. In 2014, he published a survey that found out that about 45% of cancer treatments clinical trials results could not be reproduced[76]. He launched the same year a research center on "meta-research" (METRICS) at Stanford University, analyzing the pitfalls of current medical research. In 2016, he went on with "Why Most Clinical Research Is Not Useful[77]". So be it with life science!

How about quantum research quality and reproducibility? Is it prone to the same deficiencies and biases as drugs clinical trials or some research in psychology[78]? So far, it seems faring much better and safer, at least for the part of publicly funded quantum science that is peer-reviewed and published.

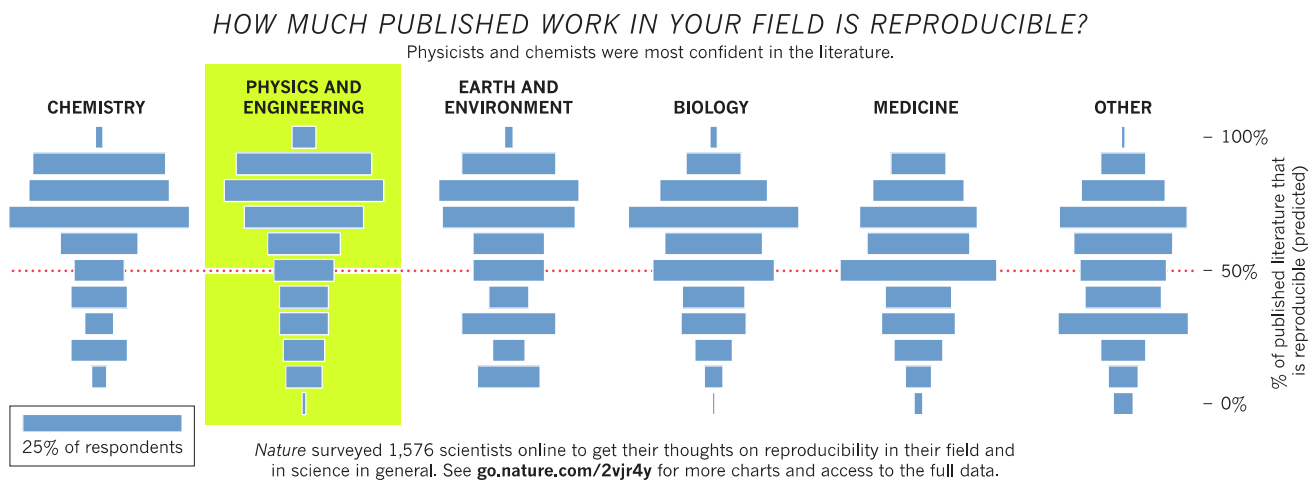

*HOW MUCH PUBLISHED WORK IN YOUR FIELD IS REPRODUCIBLE?*
Physicists and chemists were most confident in the literature.

CHEMISTRY — PHYSICS AND ENGINEERING — EARTH AND ENVIRONMENT — BIOLOGY — MEDICINE — OTHER

% of published literature that is reproducible (predicted) — 100% — 50% — 0%

25% of respondents

*Nature* surveyed 1,576 scientists online to get their thoughts on reproducibility in their field and in science in general. See go.nature.com/2vjr4y for more charts and access to the full data.

**Figure 12 : research reproducibility per science field showing it is much better in physics and engineering.**

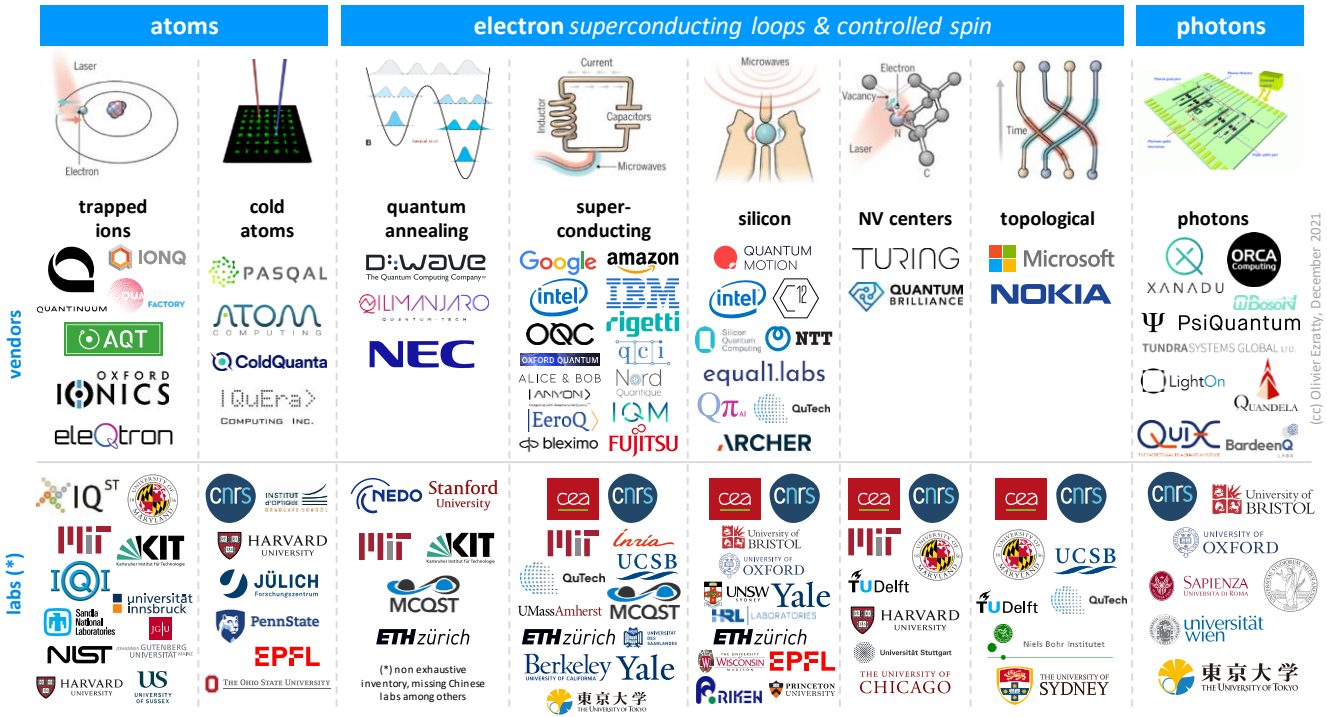

Figure 13 : key academic and vendors quantum computer players by qubit type.

Retraction Watch tracks retracted papers in peer reviewed scientific publications[79]. There, you can find only a few retracted papers in quantum physics, most of these coming from Chinese and Indian research teams. China indeed floods the science scene with tons of papers and patents, including in the quantum field. Studies show that their relative impact is much lighter than papers published in North America and Europe[80]. But it is regularly improving.

Otherwise, the most famous recent retraction was in 2021 with the withdrawal of a 2018 paper coauthored by Leo Kouwenhoven on Majorana fermions[81]. Earlier, the famous 1996 Sokal hoax concerned a fake paper created by Alan Sokal, from New York University and University College London. It was published by Social Text, an academic journal of "postmodern cultural studies". The paper created fancy links between quantum gravity and social sciences. It was created to demonstrate how a human science peer-review magazine could be fooled. So, it was not per se a quantum physics research paper.

Another survey published by Nature in 2016 did show that chemistry, physics and engineering where the fields where reproducibility was the best while life science is the worst[82] (see Figure 12). In all scientific fields, there may now be more retracted papers due to better editorial oversight, rather than to a growth of fake papers.

Most fundamental quantum physics research results are continuously improved and reproduced. For example, the best streams of continuous progress can be demonstrated with all Bell-test related quantum entanglement proofs, from Alain Aspect *et al* 1982 experiments to the 2015 loophole-free Bell test experiments which removed any remaining doubts on the 1982 findings[83].

There are however some cases when quantum research findings cannot easily be reproduced because they are very theoretical. It happens with many quantum algorithms which can only be either emulated classically at a very small scale or only performed at an even smaller scale on existing qubits. The same can said of new error correction codes.

Most of these quantum algorithms advertise speedups that are not based on using a full-stack approach including data preparation, quantum error correction and gates creation overheads. The belief is that these intermediate steps have only a small non-exponential time overhead[84].

Qubit fidelity improvements are also prone to various interpretations, particularly when they are publicized by vendors. The lack of clear standardized benchmarks makes it very difficult to compare published results, even those coming from IBM and Google which are rather well documented. We will cover that later in a set of proposals.

**Quantum technologies are diverse**

Another striking difference is the extreme technology diversity in the quantum realm. There are many competing technology routes, particularly with the types of qubits that could be used to build scalable quantum computers. And their trendiness is not very stable.

About 20 years ago, NMR (Nuclear Magnetic Resonance) qubits were considered to be key contenders. They have been nearly entirely abandoned since then. Nowadays, superconducting and trapped ions qubits seem to lead the pack. But photon and electron spins qubits loom behind and may reshuffle the quantum computing cards deck (see a table with research labs and vendors per qubit types in Figure 13). The same can be said about topological qubits which are now underwater, particularly after the 2018 Leo Kouwenhoven paper withdrawal from Nature in 2021. But many scientists are still exploring various ways of creating topological qubits, not just with Majorana fermions. We do not know yet which of these technologies could lead to some sort of scalable quantum computer exceeding the capacities of classical computers.

Although it could be perceived as slow, scientific and technology progress in quantum computing is steady. In the last 12 months, better qubits were created by many vendors like HQS (Honeywell Quantum Systems, now Quantinuum after its merger with UK's CQC) and IonQ. HQS and Google also presented the first logical qubits in 2021.

On top of that, there are several competing quantum computing paradigms: gate-based quantum computing, quantum annealing and quantum simulation. This creates a sort of fault-tolerance in the quantum computing ecosystem. Quantum simulation systems like the ones from Pasqal may bring some computing advantage earlier than imagined, way before gate-based systems. Also, although not a darling in the quantum computing world, quantum annealing and D-Wave are making regular progress and are publicizing the greatest number of potential use case for their systems.

### Quantum skeptics start to have a voice

The other difference is that, although they are a minority, skeptics have a voice. Like Serge Haroche who expressed doubts on the advent of gate-based quantum computers starting in 1996 and still now, Gil Kalai who is famous for his presentations on "why quantum computers cannot work"[85], Michael Dyakonov who made a case and a book "against quantum computers"[86] and Xavier Waintal. With colleagues from the University of Illinois and the Flatiron Institute, he did show that Google Sycamore could be emulated on a server cluster, with using some compression technique[87]. On equal feet comparisons, this quantum supremacy vanishes.

This scientific skepticism, also sometimes abusively labelled pessimism, raises interesting questions. It drives useful debates in the scientific community that is not a single unified body with regards to the prospects of fault-tolerant quantum computers. Still, more debates should be organized to confront these different views.

Overall, serious people are in charge. Most of the scientific people I know in quantum technologies are excellent professionals, even if they may have some scientific biases or are too specialized. These folks work hard and address highly complicated scientific and technology challenges. Business people who join them are often the source of overpromises and hype. Scientists feel sometimes obliged to play with these rules. These are the rule of the game in the startups funding race.

### Quantum science fact-checking requires diverse skills

The scientific and technology understanding levels required to forge some view on the credibility of scalable quantum computing are very high. It is the same in other fields like quantum telecommunications and cryptography.

It is higher than any other discipline I have investigated so far, out of the above-mentioned ones in other technology fields. On top of that, it is quite a challenge to have some sound understanding of quantum physics, quantum noise, error correction and quantum algorithm structures and real speedups altogether.

As a result, assessing the viability of quantum computing is a highly cross-disciplinary task requiring much collaboration. You can still express doubts with common sense questions[88], but these still mandate some solid scientific and engineering background.

### This is not a Schumpeterian revolution

Quantum computing is not a "jack of all trades" solution. It is not a replacement tool but more a complement to current High Performance Computers (HPC). Many, if not most of today's classical computing problems and related software are not at all relevant use cases for quantum computing. From an economic history perspective, the consequence is that quantum computing will probably not be a Schumpeterian creative-destruction case[89]. It will not entirely replace classical legacy technologies, even classical supercomputers. It will complement it. It is an incremental and not a replacement technology, like the car replacing horses or steam engines replacing animals and windmills.

You probably will not have a quantumly accelerated desktop, laptop or smartphone to run your usual digital tasks although some quantum technologies can be embedded in these devices like quantum sensors and quantum random number generators.

The only visible replacement technology generated by quantum computing is not quantum! It is Post-Quantum Cryptography (PQC), which may soon replace symmetric traditional cryptography with cryptographic key encoding that theoretically cannot be unlocked by any quantum computers. We may end-up with a Schumpeterian revolution based on the paranoia of a non-Schumpeterian revolution, that may never happen. In another field, quantum sensors may have some profound impact on some industries.

### Quantum technology sovereignty amplifies the hype

The current quantum hype is also fueled by some regions and countries who categorized quantum technologies in the sovereign technologies domain. These are technologies a modern state must either be able to create or to procure. Like with many other strategic technologies, geographical competition happens at least first between the USA and China, with Europe then sitting in between and willing to get its place under the quantum sun.

China is applying some brute force with large investments in research and massive scientific training. Their investments have however been overestimated, like the supposed $10B investment in the single lab of Hefei. Various consolidated estimates of China's public quantum investments are within the range of $2B to $4B over the last 10 years. China became the worldwide leader with quantum science publications and patents. This engendered fear in the USA, explaining the National Quantum Initiative Act launched in December 2018 and its additional funding in 2020, of $2B over 5 years with a total spending of $4B. In this matter, the US Congress is very active and in a rarely seen bipartisan way[90].

Other countries investments are also hard to compare because they are not using the same metrics: with or without existing public investments, with or without vendor investments, with or without classified defense and intelligence related investments. So, the USA $2B (federal funding, 5 years), European Union's $1B Quantum Flagship (public funding, 10 years[91]), UK's £1.2B (10 past years, public and industry funding), German 2B€ (public and industry, 5 years), French 1.8B€ (public and industry, 5 years), Dutch 625M€ (public, 6 years) and Indian $1B (public, 5 years) plans are apples and oranges.

Quantum technologies are considered as being "sovereign" for a couple reasons.

The first relates to cybersecurity. Shor's algorithm potentially threatens the cybersecurity of most of the open Internet. So, governments want both to be the first or among the few to control this new wave "cyber weapon" and simultaneously master and deploy the related "counter weapon" (post-quantum cryptography and quantum key distribution systems). The rivalry has some similarities with what has happened since WWII with nuclear weapons. There is a narrow, but unfortunately expanding club of countries able to manufacture it, North Korea being the last one. Never used since Nagasaki, it has been a powerful tool for regimes as a deterrent against potential adversaries. The difference here is that the counter-weapon protecting systems against quantum computers will be available way before these can break RSA-based public keys.

The second is the European Union is traumatized by its dependence on the USA and Asia for microprocessor design and manufacturing, for key Internet services (with the so-called GAFA: Google, Apple, Facebook and Amazon) and, lately, with cloud infrastructures being dominated as well by US vendors, at least in the Western world. Everywhere, even in the USA, the fear of the emergence of technology monopolies is also strong. Europe is eager to get its share with quantum technologies, leveraging a potential opportunity to level the playing field with a new technology wave, having missed most previous digital waves. The European Union and its member states are collectively investing about $5B of public funding to fuel the quantum race.

The third is "dual usages", where quantum technologies can be applied in the civil and commercial spaces and in the military/intelligence. As with any other technology like supercomputing, developed countries want to secure their ability to develop and/or acquire technologies applicable to these non-civil use cases. In November 2021, the US State Department added China's Hefei National Laboratory for Physical Sciences at Microscale and QuantumCTek to a black-list of entities suspected to aid the Chinese army in the development of advanced radars, undersea sensors, encryption breaking tools and unbreakable encryption[92]. Exports controls also deal with enabling technologies like low-temperature cryogenic systems.

Securing special raw materials and isotopes is another strategic concern (rare earths/lanthanides, $Si^{28}$, $He^3$, etc.) as well as access to some critical enabling technologies. Dual use cases also raise the stakes with the need for a responsible innovation approach as discussed later in the Towards responsible quantum innovation section and the potential dilemma it could create with the involved scientists. Even when it can be argued that many use cases for quantum technologies in the military are defensive and not actual weapons per se.

How does all of this relate to the quantum hype? It is a key driver of developed countries' funding for both quantum fundamental research and technology developments, including startups public funding, either directly (grants, loans, capital) or indirectly (procurement). The phenomenon is however not specific to quantum technologies. It was already in play in many other sectors like with artificial intelligence but it looks like the ratio of publicly funded research to commercial vendors investments (including venture capital) is one of the highest compared to other digital technologies.

Furthermore, national security concerns tend to exaggerate threats, thus fueling the hype[93]. Beyond the hunt for the creation of scalable quantum computers able to break symmetric public keys infrastructures, quantum sensing also has its fair share of hype in the military like with so-called quantum radars, whose feasibility is far from being proven despite China's claims. These cannot be taken at face value[94]. It circles back to the complexity of fact-checking science and technology news in the quantum realm.

Finally, when a technology is becoming "sovereign", particularly with national security use cases, its hype has potential negative side effects. It is prone to propaganda, the strongest coming from China and its various technology premieres that are even more exaggerated than the ones coming from the USA like Google's Sycamore supremacy. It also tends to limit international cooperation and as a result reduces the capacity to fulfill expectations that would require a coordinated worldwide mobilization. Thus, there is a need to differentiate what belongs to worldwide shareable fundamental research and where it becomes strategic and not shareable, especially given these concerns are also emerging as quantum technologies enter the commercial vendors space.

## QUANTUM HYPE MITIGATION APPROACHES

Technology hypes are inevitable. But there are ways to mitigate their most undesirable effects like a sluggishness to deliver results or a decrease of trust in scientists, engendering some winter that would shrink or kill government, investor and corporate R&D funding sources.

There are some solutions to explicitly investigate. It is a matter of general conduct for the whole quantum technologies community but also some specific actions.

### Explain the progress

Although scalable quantum computers have not yet seen the light of day, quantum science is making a lot of progress. Qubit fidelities are improving in many of their implementations like superconducting or trapped-ions. 2021 saw interesting advances with the creation of the first logical (error-corrected) qubits with Google (superconducting) and Honeywell Quantum Systems (trapped ions). There were also many documented advances in cryo-electronics like the presentation of the HorseRidge 2 cryo-CMOS component from Intel[95] or announcements made by SeeQC with superconducting qubit control electronics and their partnership with OQC in the UK.

What is amazing in the quantum computing space is the sheer diversity of technologies being developed. NV centers made kind of a comeback in the quantum computing space with announcements coming from Quantum Brilliance, a German-Australian startup, even though it was a bit oversold with its "room temperature" operations. There were also some breakthroughs with photon qubits, particularly with the first parametrized (and soon programmable) boson sampling setup achieved by a Chinese lab led by the country quantum "czar" Jianwei Pan.

At last, even though it requires investigation, some research labs and vendors described scale-out approaches, that would be based on quantumly interconnecting several quantum processing units (IonQ, Rigetti, IBM), with photons and qubits-photons conversion mechanisms. A few startups were even created in this field, including QPhoX, in the Netherlands, Next Generation Quantum in the USA and PhotoniQ in Israel.

There were also some advances in the complicated realm of quantum memory, a technology that will be strategic to enable quantum computing, particularly with quantum machine learning and some oracle-based algorithms.

Progress reviews could be created and published by government agencies and research laboratories, and preferably with independent expert bodies. There are many review papers being published on specific technologies. They are often hard to read for non-specialists, given they refer to dense bibliographies and use some complex wording. An intermediate level of explanations would also be nice. This will mandate some investment from the research community and their governing organizations[96].

Some additional work could be done to position the technology readiness level (TRL) of key quantum technologies and how it is evolving over time, as well as common definitions for key terms like quantum advantage and supremacies[97]. This is particularly true to the ever growing list of exotic qubit types that are being created by research labs at very low TRL, like with silicon carbide or molecular qubits.

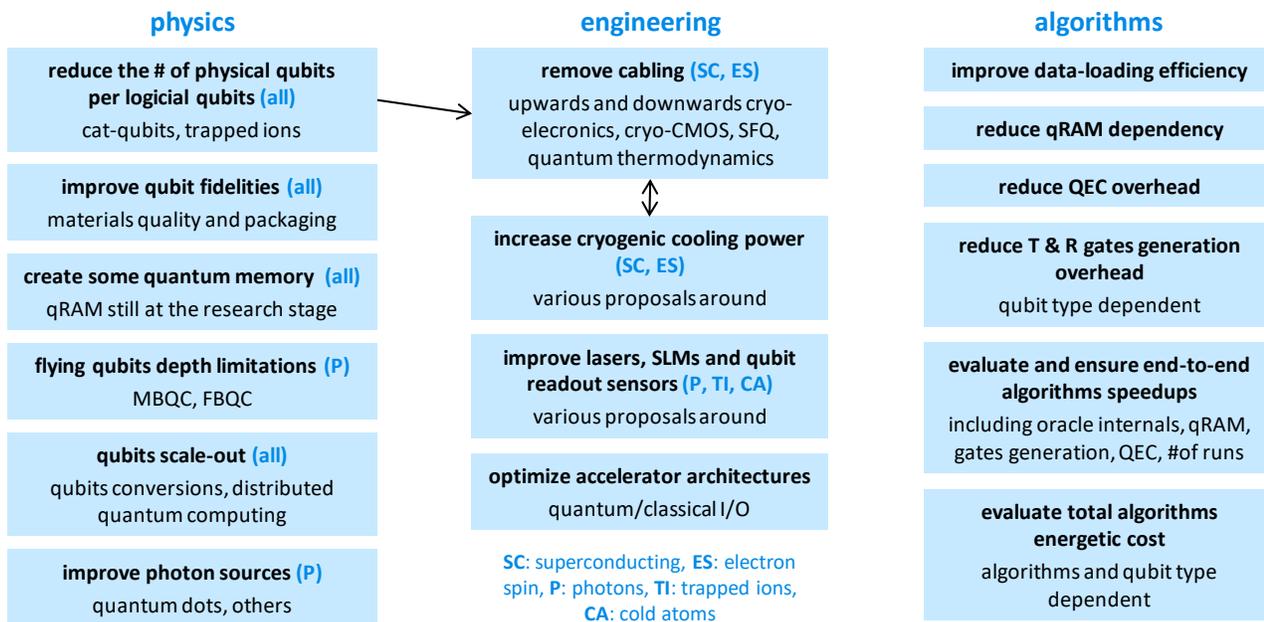

| physics | engineering | algorithms |
|---|---|---|
| **reduce the # of physical qubits per logicial qubits (all)** cat-qubits, trapped ions | **remove cabling (SC, ES)** upwards and downwards cryo-elecronics, cryo-CMOS, SFQ, quantum thermodynamics | **improve data-loading efficiency** |
| **improve qubit fidelities (all)** materials quality and packaging | **increase cryogenic cooling power (SC, ES)** various proposals around | **reduce qRAM dependency** |
| **create some quantum memory (all)** qRAM still at the research stage | **improve lasers, SLMs and qubit readout sensors (P, TI, CA)** various proposals around | **reduce QEC overhead** |
| **flying qubits depth limitations (P)** MBQC, FBQC | **optimize accelerator architectures** quantum/classical I/O | **reduce T & R gates generation overhead** qubit type dependent |
| **qubits scale-out (all)** qubits conversions, distributed quantum computing | **SC**: superconducting, **ES**: electron spin, **P**: photons, **TI**: trapped ions, **CA**: cold atoms | **evaluate and ensure end-to-end algorithms speedups** including oracle internals, qRAM, gates generation, QEC, #of runs |
| **improve photon sources (P)** quantum dots, others | | **evaluate total algorithms energetic cost** algorithms and qubit type dependent |

Figure 14 : map of some quantum computing scientific and technology challenges.

## Map the challenges

Quantum computing critics must be listened to. It requires some work in mapping the challenges ahead (see Figure 14 with a tentative challenges map of my own). It is a mix of theoretical, scientific and engineering challenges, the best example being the links between quantum thermodynamics and cryo-electronics. But don't count on the magic wand of engineering to solve quantum computing' daunting scientific challenges[98].

A lot must also be done with quantum algorithms and quantum software engineering. It is hard to define the real speedups of quantum algorithms when using a full-stack approach including data preparation, gates preparation, error correction and multiple runs. Data-preparation is a long process for some algorithms.

That is why we can position quantum computing benefits to sit at the crossroads of complex calculations and relatively small data sets. For a very long time, quantum computing will not be adapted to "big data" applications.

## Consolidate transversal projects

Due to its diversity, quantum technology efforts are far from being concentrated or consolidated, even in the USA and China. The efforts are scattered in many laboratories and vendors, since many different technology avenues are being investigated. Many physicists argue that it is way too early to launch consolidated initiatives given the lack of maturity of qubits physics. Above all, they want to avoid governments mandating qubit technology choices too early, reminding them of the fate of NMR qubits which were fashionable 20 years ago and are not anymore.

To make a comparison with the Manhattan project, it is like if the USA had worked on 12 different atomic bombs instead of two (uranium and plutonium based, although classical explosives triggering nuclear critical mass required some iterations in their design). Quantum physicists gather in highly specialized conferences (virtually or IRL) while a new breed of "quantum business" conferences give voice to (sponsor) vendors, analysts and some corporate users with usually overly optimistic statements. Next comes the notion of quantum engineering, a cross-discipline approach mixing sciences (physics, mathematics, thermodynamics, etc.) and engineering (systems design, electronics, cryogeny, software engineering, etc.) to solve complex design problems. It can also create better serendipity, spurring unexpected scientific and technological progress in adjacent domains.

When you look at the shape and form of government quantum plans, they are most of the times piecemeal. Some have created technology hubs, like in the UK and the USA. The large Hefei research lab in China is in the same vein. So is the CQT in Singapore, created in 2007. But it does not necessarily translate into a joint physics-engineering coordinated effort.

There are some domains where it would make sense. I see at least a couple ones: in cryo-electronics, with scale-out architectures connecting several quantum processing units together through photonic links and with quantum technology energetics. Can we ensure quantum computers will bring some energetic advantage and not see their scalability be blocked by energetic constraints[99]?

These are perfect cases where multidisciplinary science and engineering are required. Hardware and software teams would also benefit from working more closely together.

At some point, creating a fault-tolerant quantum computer will look more like building a passenger aircraft, a nuclear fusion reactor ala ITER instead of a creating a small-scale physics experiment. There, much more international coordination and collaboration will be required to assemble all of it in a consistent manner.

## Revisit benchmarking

Creating rational ways of comparing quantum computers performance from one vendor to another and across time is fundamental to set the stage. IBM's Quantum Volume was a first attempt, launched in 2017 and adopted in 2020 by Honeywell and IonQ. It is however a bit confusing, particularly given quantum volume cannot be computed in a quantum supremacy or advantage regimes since it requires doing some emulation on classical hardware.

Many other benchmarks have been proposed, by the Oak Ridge National Laboratory for chemical simulation[100], by QuSoft, the University of Cambridge and Caltech to measure the performance of universal quantum computers in a hardware-agnostic way with six structured circuits tests[101] and by researchers from Princeton, HQS, QCI, IonQ, D-Wave and Sandia Labs as a collaboration driven by the Quantum Economic Development Consortium (QED-C) in the USA, and with a series of application oriented tests[102].

The IEEE also launched several benchmarking initiative on its own[103]. Atos's Q-Score proposes on its side a simple metric defining the maximum size of some optimization problem that can be solved on a given system[104].

Having vendors accepting the implementation of third-party benchmark tools is important to create some form of trust. These benchmarks will lead to the creation of quantum computers rankings like the Top500 HPC.

Without broadly accepted benchmarks and rankings, wild announcements of XYZ company with W qubits will continue to pop-up and that are hard to fact-check. The confusion may be amplified by some hardware vendors creating hardware specialized for specific tasks before the potential advent of universal quantum computers able to solve any, and all problems a quantum computer could address.

At some point, we will also need to create energetics oriented benchmarks and rankings, to ensure quantum computers scaling is not done at the cost of astronomical energy costs. Some innovative benchmarking techniques could also be created to compare the three main quantum computing paradigms (annealing, simulation, gate-based). It would make sense given a large portion of quantum algorithms like chemical simulations and complex optimizations can run on these three architectures.

Finally, while numerous new quantum algorithms are published, their real speedups are difficult to assess with existing hardware. A clear mapping of existing proof of concepts and their various operational benefits should be made available by vendors and consolidated by independent third parties (universities, scientific foundations, industry consortiums).

### Invest in education

Quantum computing is probably one of the most complicated domains to assess for information technology specialists, and of course, by the overall digital ecosystem and the general public. On many occasions, the topic is covered in short formats whether online, in events or videos. What it is to be quantum remains largely a mystery for the general public and even for most engineers and information technology specialists.

One key challenge is to be concise and precise when describing the whereabouts of quantum computing. Educating the IT workforce will become a priority to create skills around quantum hardware, software and services.

We need educational programs for students in undergraduate, graduate and doctoral schools and for existing IT professionals, particularly developers, with continuous and/or self-training. Also, the general public education should not be forgotten given the audience is highly vulnerable to vendor hype and fake science news[105].

The need will however slowly grow as the technology matures. Investors and policy makers will also need to be educated. Specialized media will need to improve their fact-checking capabilities and more seriously pinpoint vendor abuses', when marketing fluff entirely occludes science[106]. Research labs and vendors will also need to improve their scientist visibility.

Only a few quantum scientists are stars in their own country. I heard it was the case of Carlo Rovelli in Italy. Are John Preskill and Scott Aaronson visible in US mass media? Not so sure. We need to have many Carl Sagan of quantum computing.

### Towards responsible quantum innovation

Quantum ethics may be perceived as being very distant from hype mitigation. It is not. As Victor Galitski states in his already mentioned paper, the ethics of science communication is at stake in the current quantum hype.

Overall, science and technology ethics should not be an afterthought, studied only when technology is ready to flood the market. It starts with studying the potential ethical and societal issues created by quantum technologies with Technology Assessment methodologies and can then lead to Responsible Research and Innovation approaches that take into account ethics, societal values and imperatives early on, like fighting climate change[107]. It tries to anticipate the unintended consequences of innovation before it occurs, even though innovation processes are inherently messy and unpredictable and when the perspectives of quantum technologies are polluted by off-the-mark wild claims[108].

It can also lead to create new regulations such as on privacy. It should also include the matter of resources like raw materials and energy. These are unavoidable constraints, particularly as materials are finite and electricity production has still a significant $CO_2$ emission cost.

It involves the organization of the dialog between many stakeholders like researchers in STEM[109] as well as in human sciences, politics, vendors and NGOs. It must preferably be an international approach encompassing all cultures and referential models, bringing the ability to understand a greater breadth of problems and increase the capability to create related solutions.

Quantum ethics has been a research investigation subject for about ten years with science philosophers, particularly in the Netherlands, the UK, Australia and France.

Australia was the first country where a quantum ethics initiative was formerly initiated. Early on, in 2019, CSIRO, the Australian scientific research agency, mentioned the need to explore and address any unknown ethical, social or environmental risks that may arise with the next generation of quantum technologies[110]. It was followed in 2021 by a white paper published by Elija Perrier from the Centre for Quantum Software and Information at the Sydney University of Technology[111]. The paper covers ethical quantum computation and asks many ethics related questions that could be asked for any kind of classical computing. They touch on the complicated question of quantum algorithms auditing and mention the need for some Quantum Fair Machine Learning (QFML). They even go as far as asking whether quantum interferences implemented in quantum algorithms are ethical in nature. They also cover privacy and cryptography matters. Is Shor going to kill our private life? How could some differential privacy be implemented with quantum computing? Other topics involved distributional ethics and fair distribution which are classical economical questions arising with any new technology. Lastly, they wonder about the impact of quantum simulations and whether it could be implemented to simulate people's personal behavior.

In the UK, quantum responsible innovation is investigated by researchers from the Universities of Oxford and Edinburgh[112]. They are assessing the challenges to embed responsible innovation in the UK quantum programme. They are looking, like Australia, at the impact of quantum machine learning, even though its real capacity to handle large volumes of data is still questionable. They are also looking at the various use case of quantum technologies, including quantum sensing, in the Military[113].

Ethical quantum computing became a topic promoted by The Quantum Daily, a US-based media platform, starting in December 2020. They released a short video documentary trying to explain what quantum computing is and the related ethical issues involved with researchers like John Martinis and entrepreneurs like Ilana Wisby[114]. They highlight the need for democratizing quantum technology skills, mention the risks on privacy and security and the need to address quantum AI bias. They also pinpoint the "Hype-Fear-Disappointment Cycle" and recommend to setting realistic expectations to avoid triggering fears and biases.

In the Netherlands, the government 615M€ initiative launched in April 2021 includes a 20M€ plan on quantum ethics and societal impact research run out of the Living Lab Quantum and Society spun out of Quantum Delta NL, the foundation established to run the Netherlands quantum program. They also create ethical, legal and societal standards for quantum technologies and their applications.

The World Economic Forum launched its Quantum Computing Governance initiative in February 2021 and published a set of governance principles in January 2022[115]. It wants to standardize an ethical framework enabling the responsible design and adoption of quantum computing. They ask the ever-lasting question: will the public trust technologies which they cannot understand and whose results they cannot verify (as if they verify it with existing digital technologies...). The WEF advocates the use of preemptive involvement in technology design to make sure ethical issues are addressed as early as possible. With that, they are assembling a "*global multistakeholder community of experts from across public sector, private sector, academia and civil society to formulate principles and create a broader ethical framework for responsible and purpose-driven design and adoption of quantum computing technologies to drive positive outcomes for society*". They will frame the conversation, drive quantum ethical issues awareness, study quantum related risks, design quantum computing ethics principles and framework and test it with some case studies.

In Canada, Q4Climate is an initiative and think tank encouraging the positive use of quantum technologies in climate research. For example, it explains how some quantum chemistry algorithms could potentially solve some environmental problems. It is comprised of researchers coming from the Institut Quantique, the University of Waterloo, Zapata and others.

In Germany, PushQuantum: Climate is a new initiative adding to this research. It acknowledges the importance of increased dialogue around the topic and also between climate and quantum communities. They have found the climate crisis also serves as a strong basis around which to take a broader consideration of the role of technology in society.

Other issues can emerge like a potential rebound effect due to an expansion of quantum computing use, creating yet another significant source of energy consumption. This is difficult to forecast at this point in time given the immaturity of quantum computing and the uncertainty about its large-scale potential. Many mass-market use-cases for quantum computing like automated road traffic optimization and personal medicine are indeed highly dependent on the availability of relatively large-scale quantum computers.

In the USA, some spare initiatives were launched by academics, a while ago, by Scott Aaronson and more recently, by Chris Hoofnagle from Berkeley, who even built some scenarios in case of a quantum computing bubble burst[116]. And there is still room for other initiatives!

## CONCLUSION

We have seen that the quantum hype has many differentiated aspects compared to past and current technology hypes, the main ones being its technology diversity and the complexity of evaluating its scientific advancements and roadmaps. Useful quantum computing is probably bound to be a long lasting quest, spanning several additional decades. Patience will be a virtue for all stakeholders, particularly governments and investors.

Among quantum technologies, quantum computing has a rather low technology readiness level (TRL) with a high uncertainty on the outcome of scalable quantum computers. Research in this domain is spread among academic institutions and small to large vendors.

As a result, an open fact-driven and science-driven approach must be pursued to evaluate progresses and roadmaps. It will mandate a greater visibility and engagement of quantum scientists in media but also a more transparent behavior from scientists made entrepreneurs.

More collaboration and coordination to make scalable computing a reality could also be beneficial to all parties and countries, a bit like what is currently done around nuclear fusion.

Mitigating the hype also require significant efforts in professional audiences as well as public education. Finally, responsible innovation and ethics of quantum technologies should be investigated and implemented as early as possible to ensure society and policy-makers trust in quantum technologies[117].

The author thanks Alain Aspect, Alexia Auffèves, Christophe Jurczak, Cyril Allouche, Emily Haworth, Jonathan Attia, Michel Kurek, Neil Abroug, Thierry Chanelière and Tristan Meunier for their feedback and the many other quantum scientists, entrepreneurs and stakeholders with whom I discussed about the topic.

v2 incorrectly stated (with an incorrect source) that Scott Aaronson was behind the Quantum Bullshit Detector 2019-2021 Twitter account.

## Table of figures

## Sources and notes

[14] You may need to spend some time explaining why quantum teleportation will not enable Star Trek like teleportation of whole bodies. Also, to bring some nuance when hearing the classical cliché arguments on exponential innovations, comparisons with other technologies, with past misunderstood scientists and the likes.

[15] See https://www.gartner.com/en/research/methodologies/gartner-hype-cycle and https://en.wikipedia.org/wiki/Gartner_hype_cycle. And Scrutinizing Gartner's Hype Cycle Approach by Martin Steinert and Larry Leifer, Stanford University, 2010 (14 pages) which analyzes critically the Gartner model.

[16] See Technology Hype Curve 2 by Richard Veryard, July 2009.

[17] See 8 Lessons from 20 Years of Hype Cycles by Michael Mullany, December 2016 and Cheap shots at the Gartner Hype Curve by Jorge Aranda, October 2006.

[18] See The Dynamics of Hype by Richard Veryard, February 2013.

[19] See Technology Hype Curve by Richard Veryard, September 2005.

[20] See Technology Hype Curve 3 by Richard Veryard, August 2009.

[21] See Crossing the Chasm, Geoffrey Moore, 1991, 2014 (288 pages) and Inside the Tornado, Geoffrey Moore, 1995 (272 pages). And Crossing the Chasm: Technology adoption lifecycle by Mithun Sridharan, not dated.

[22] NISQ stands for Noisy Intermediate Scale Quantum computers, quantum computers using noisy non-corrected qubits in the range of 50 to a couple hundreds, and which are supposed to be useful for certain tasks, if not provide a quantum computing advantage, meaning the capacity the run these tasks more efficiently than on classical computer (in time, cost and/or spent energy).

[23] See Disruptive Technologies: Catching the Wave by Joseph L. Bower and Clayton M. Christensen, Harvard Business Review, 1995 and What Is Disruptive Innovation? Twenty years after the introduction of the theory, we revisit what it does—and doesn't—explain. by Clayton M. Christensen, Michael E. Raynor, and Rory McDonald, Harvard Business Review, 2015.

[24] See Is it time to hire a chief metaverse officer? by Maghan McDowell, Vogue Business, October 2021.

[25] See China's top regulators ban crypto trading and mining, sending bitcoin tumbling by Alun John and Samuel Shen and Tom Wilson, Reuters, September 2021.

[26] See Blockchain Startups Raised over $4 Billion in VC Funding in Q2 2021 by Anthonia Isichei, July 2021.

[27] See The architecture of a web 3.0 application by Preethi Kasireddy, 2021 and Web3 is Bullshit by Stephen Diel, 2021. This author makes an interesting analogy between the crypto craze and quantum fake sciences in Decentralized Woo Hoo, 2020.

[28] Barbados even opened an embassy in a Metaverse! See Barbados to Become First Sovereign Nation With an Embassy in the Metaverse by Andrew Thurman, November 2021.

[29] See https://www.energystartups.org/top/fusion-energy/. And I don't mention the so-called cold fusion frenzy.

[30] See Nuclear fusion has attracted more than $2 billion in VC and private funding by Eric Wesoff, June 2021.

[31] I have compared the trendiness of various terms with Google Trends which tracks the relative number of times Google Search users are looking for these. It gives a good indication of the awareness of these terms evolution over time. The charts below follow this over a period of five years. It shows that the quantum hype, if any, is relatively stable over time, with a peak in October 2019 linked to Google's supremacy announcement.

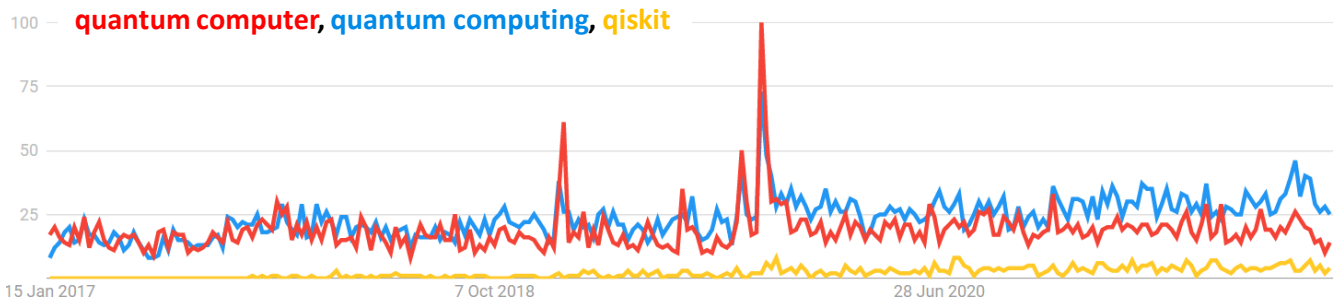

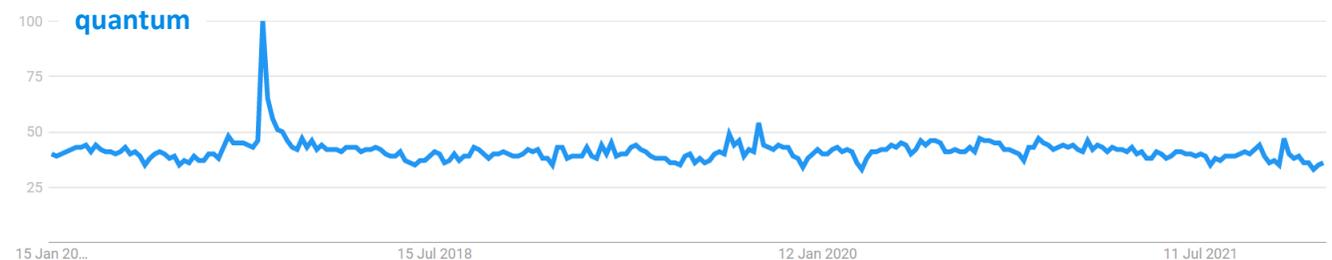

Next is a comparison with other terms, first with non-scientific hypes and then with other scientific hypes. Clearly, the quantum hype noise is nothing compared with Blockchain, NFTs and metaverses awareness, and the Bitcoin is much higher than these and not charted here. In science, quantum computing awareness is below nuclear fusion but with a similar order of magnitude, and above narrower life-science hypes like DNA sequencing and precision medicine. But quantum physics is above nuclear fusion.

If you boil in mRNA vaccines or covid related science, these are several orders of magnitude more searched than quantum computers and nuclear fusion. These curves show that, at least quantitatively, the quantum computing hype does not follow a Gaussian trend like in Gartner's hype curve model. However, if we were analyzing quantitatively and qualitatively the quantum hype reaching corporations decision makers, the picture would maybe be different.

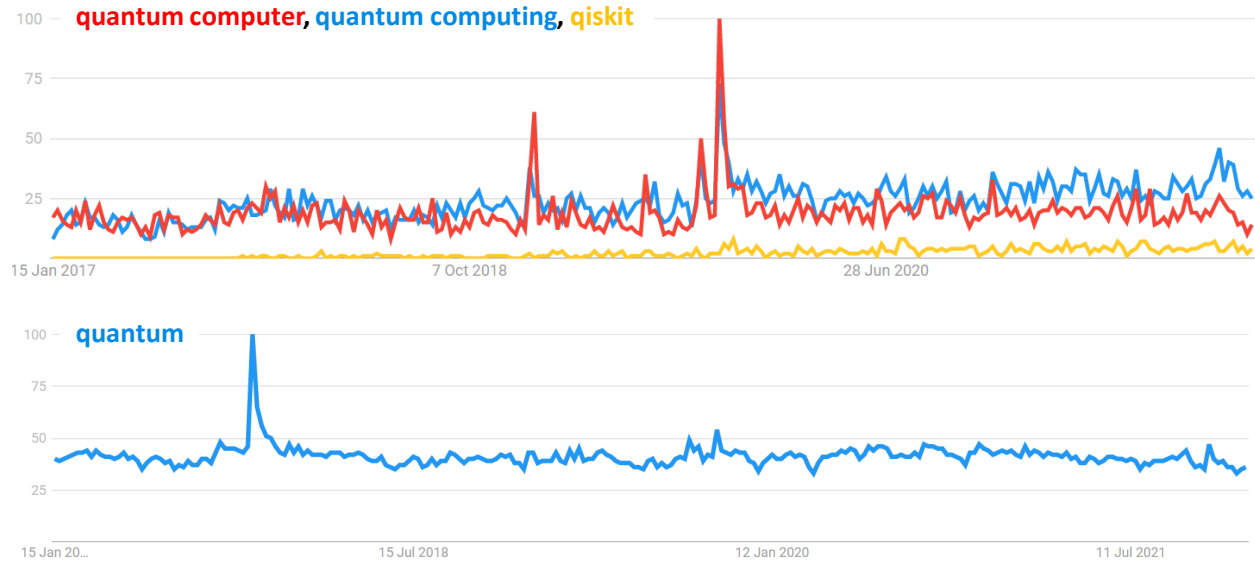

Google Trends, worldwide, January 11st, 2022

[32] Peter Shor also created his less talked about discrete log (dlog) quantum algorithm that could be used to break elliptical curves based encryption.

[33] The origins of the first Internet bubble are multiple. The stock market had crazy high valuations for large technology corporations and telecom operators, driven by low interest rates. It made investments in venture capital an attractive asset category. This phenomenon is the same today, that explains why so much money is poured into startups ($621B worldwide in 2021). In 2000-2001, the Internet startups ecosystem couldn't yet strive since the equipment rate of Internet devices, mostly personal computers, was then rather low in households (13% in Europe in 2000) and Internet bandwidth was limited by dial-up modems speeds of 56 Kbits/s. Then, in the mid-2000s, broadband started to be deployed, providing speeds of several Mbits/s. At last, starting in the late 2000s, mobile Internet, 3G cellular networks and smartphones contributed to growing the Internet overall market. All-in-all, these was a dephasing between the first Internet financial bubble and the infrastructure sockets (personal computers and telecoms) available to grow usages. Finally, Internet waves were more dependent on the telecom equivalent of Moore's law than its application on microprocessors. Meanwhile, many key Internet software sockets were developed in the early years of Internet 1.0 (Apache, PHP, MySQL, Java, JavaScript, etc) and are still in use today.

[34] See [Beijing and Vienna have a quantum conversation](#) by Hamish Johnston, Physics World, June 2021.

[35] Google's quantum supremacy's related propaganda continues to resonate over two years after it started. Particularly with the false equivalency of 2 (quantum) minutes vs 10.000 (classical) years computation time comparison for some random computing that has no input data and generate a good result only 0.2% of the time.

[36] See [Quantum Computational Advantage via 60-Qubit 24-Cycle Random Circuit Sampling](#) by Qingling Zhu, Jian-Wei Pan et al, September 2021 (15 pages).

[37] See [Phase-Programmable Gaussian Boson Sampling Using Stimulated Squeezed Light](#) by Han-Sen Zhong, Chao-Yang Lu, Jian-Wei Pan et al, June 2021 (9 pages).

[38] China research however have reached records with photon based qubits using the boson sampling technique. This technique is not yet usable for programming a problem with some input data. See [Phase-Programmable Gaussian Boson Sampling Using Stimulated Squeezed Light](#) by Han-Sen Zhong, Chao-Yang Lu, Jian-Wei Pan et al, June 2021 (9 pages).

[39] Even in the relatively sober [Quantum Computing for Business Leaders](#) by Jonathan Ruane, Andrew McAfee, and William D. Oliver, Harvard Business Review, 2022, the use cases list is quite optimistic with, among others, faster and more accurate Netflix-like recommendations, a future Pixar movie made with quantum generated artificial content using generative adversarial networks and structured search using Grover algorithm at large scale (without mentioning its very poor quadratic speedup).

[40] See [Forecasting the Weather Using Quantum Computers](#) by 1Qbit, 2017. The paper references [CES 2019: IBM unveils weather forecasting system, commercial quantum computer](#) by Abrar Al-Heeti, January 2019, which covers two entirely unrelated announcements from IBM, one on weather forecasting using classical computing and another, related to their Q System One, both introduced at CES 2019. Off we go, and quantum computing makes weather forecasts!

[41] See [IBM CEO: Quantum computing will take off 'like a rocket ship' this decade](#) by Mark Sullivan, FastCompany, September 2021.

[42] See [https://www.classiq.io/insights/2021-survey-part-1](https://www.classiq.io/insights/2021-survey-part-1).

[43] See this fear being played out in [Report: 74% of Executives Warn Either Adopt Quantum Soon, or Risk Falling Behind Forever](#) by Matt Swayne, The Quantum Daily, January 2022 and [Report: 69% of enterprises embrace quantum computing](#) in VentureBeat, January 2022. It deals with the first annual report on enterprise quantum computing adoption commissioned by Zapata Computing. Interestingly, data analytics is positioned as a key use case for quantum computing which is really far-fetched given quantum computers won't probably be competitive for processing big-data. And 41% of the polled enterprises expect a business competitive advantage coming from their adoption of quantum computing within the next two years, when no one really knows if and when we'll have a real "quantum advantage" with quantum computers in that timeframe. This is typical of a technology hype "peak of expectations"

period. See also [Quantum Computing — New Paradigm or False Dawn?](#) by Stuart Robertson and David-Jordan from L.E.K., a global consulting firm, January 2022, which rehashes the same quantum hype messages and paints a misleading picture of the progress with quantum computer's power (see their Figure 3, Rigetti never released their 128 qubits system, neither Google's 72 qubits, etc)..

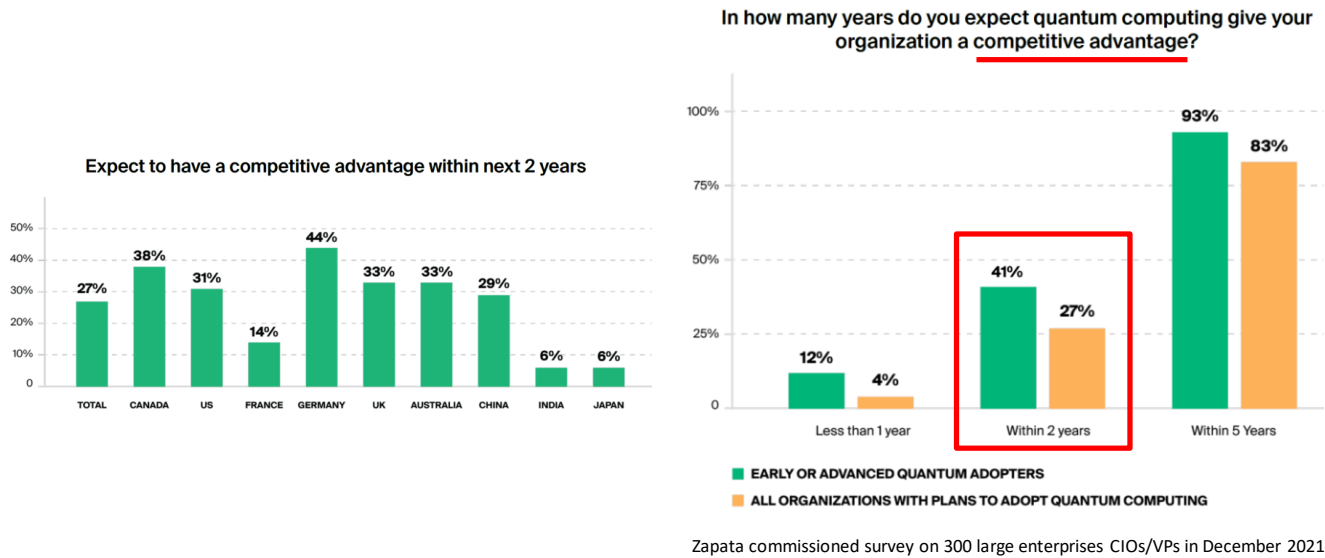

Zapata commissioned survey on 300 large enterprises CIOs/VPs in December 2021

[44] This chart comes from my ebook "Understanding Quantum Technologies" published in September 2021. Free download in PDF and available in paperback edition on [Amazon](#). See [https://www.oezratty.net/wordpress/2021/understanding-quantum-technologies-2021/](#).

[45] See [The Rigetti 128-qubit chip and what it means for quantum](#) by Rigetti, August 2018. In 2021, Rigetti announced a 80 or 40+40 qubit system which make things confusing. Indeed, 40+40 qubits have much less computing capacity than 80 qubits unless the two 40 qubit sets are entangled at some level.

[46] A variation of the Quantum Volume denominated logarithmic Quantum Volume by IonQ corresponds to a number of qubits. A $\log_2(QV)$ is the number of qubits corresponding to a Quantum Volume QV. This is the number of qubits that pass a randomized benchmarking test generating a correct result in at least 2/3 of the cases.

[47] See [Validating quantum computers using randomized model circuits](#) by Andrew W. Cross et al, October 2019 (12 pages).

[48] IBM's quantum volume doubling every year still happens with their 27 qubits processors. Their quantum volume is in practice having ups and downs. Their latest 27 qubits system reaches a QV of 128. Their 65 qubits have a QV of 32 and their 127 qubits started at 32 in December 2021 and were at 64 in January 2022.

[49] See [VR/AR to reach $162 billion in worldwide revenues by 2020 - IDC](#), August 2016.

[50] See [Virtual Reality Market Size, Share & Trends Analysis Report (2021 - 2028)](#), March 2021.

[51] See [Quantum Computer Market Headed to $830M in 2024](#) by John Russell, HPC Wire, September 2021.

[52] See [The Worldwide Quantum Technology Industry will Reach $31.57 Billion by 2026 - North America to be the Biggest Region](#), Research and Markets, May 2021. I bet it won't!

[53] See [Quantum computing will be the smartphone of the 2020s, says Bank of America strategist](#) by Chris Matthews, December 2019.

[54] See [Quantum computing will be worth $1 trillion by 2035, according to McKinsey](#), March 2020.

[55] In 2013, Cisco predicted that the value created by the Internet of things would reach $14T in [Embracing the Internet of Everything To Capture Your Share of $14.4 Trillion](#), 2013 (18 pages). See also [Green Hydrogen The next transformational driver of the Utilities industry](#) by Goldman Sachs, 2020 (60 pages), which predicts the hydrogen "addressable" market will reach $10T by 2030, so about 8% of the worldwide GDP.

[56] Look for example at [Bonus Report #1: "Quantum Lotto: The Quantum Computing Company Shooting for the Moon"](#). This report that is supposed to uncover the name of one quantum company to invest in is a freebie for a $2K subscription to an investor newsletter. The web site is a consolidation of all the quantum hype available.

[57] IDC quarterly 2020 server market estimates: [Q1](#), [Q2](#), [Q3](#) and [Q4](#) with respectively $18,6B, $18,7B, $22,6B and $25,8B.

[58] See [https://www.statista.com/outlook/tmo/software/worldwide](#).

[59] See [What Happens When 'If' Turns to 'When' in Quantum Computing](#), BCG, July 2021 (20 pages).

[60] See [Quantum Computing as a Service Market Sizing - How we did it](#), The Quantum Daily, August 2021.

[61] See [QC ethics and hype: the call is coming from inside the house](#) by Scott Aaronson, March 2021 and [Will Quantum Computing Ever Live Up to Its Hype?](#) By John Horgan, Scientific American, April 2021.

[62] Other claims are questionable like their large quantum volume. IonQ advertised having reached a quantum volume of 4 million, corresponding to using 22 qubits and a safe depth of 22 series of quantum gates. But they didn't publish a documented benchmark with the related 32 bits quantum system. Users can only test their 11 qubits system running on Amazon, Microsoft and Google's clouds.

[63] See for example [High stability cryogenic system for quantum computing with compact packaged ion traps](#), by Robert F. Spivey et al, including four authors working at IonQ, August 2021 (12 pages).

[88] One example when some vendor announces to have developed a quantum processor with a very large number of qubits: are these entangled?

[89] Creative-destruction is a concept created by the economist Joseph Schumpeter where a given innovation destroys some existing economical agents and replace them with new ones. The most famous examples are cars replacing horses and their related ecosystems. To some extent, personal computers (and their users) replaced secretaries.

[90] See The Great Rivalry: China vs. the U.S. in the 21st Century by Graham Allison, December 2021 (52 pages).

[91] See Scientists and citizens: getting to quantum technologies by David P. DiVincenzo, 2017 (5 pages) which explores what quantum technologies can deliver when analyzing the promises presented in the European Quantum Manifesto in 2016. This is also done in Quantum hocus-pocus by Karl Svozil, 2016 (6 pages), which adopts a more critical view of the overpromises from the Manifesto.

[92] See Addition of Entities and Revision of Entries on the Entity List; and Addition of Entity to the Military End-User (MEU) List, November 2021.

[93] See Quantum technology hype and national security by Frank L Smith, April 2020 (18 pages).

[94] See China's claim of developing "quantum radar" for detecting stealth planes: beyond skepticism by Ashish Gupta, 2016 (4 pages) and The US and China are in a quantum arms race that will transform warfare by Martin Giles, MIT Technology Review, January 2019. Some scientists, "under the radar", find that quantum radars features are overstated.

[95] See A Fully Integrated Cryo-CMOS SoC for Qubit Control in Quantum Computers Capable of State Manipulation, Readout and High-Speed Gate Pulsing of Spin Qubits in Intel 22nm FFL FinFET Technology by J-S. Park et al, February 2021 and 41 slides.

[96] See Quantum Computing 2022 by James D. Whitfield et al, January 2022 (13 pages) ambitions to demystify the state of the art of quantum computing. However, it is a very complex read for the non-initiated and deals mostly with algorithms but not much with hardware and its state of the art and challenges.

[97] The QTRL (quantum technology readiness level) scale created by Kristel Michielsen can be used here.

[98] See Quantum computing as a field is obvious bullshit by Scott Locklin, January 2019. His main argument, and a valid one, is that creating a usable quantum computer is not just an engineering challenge but a tough science one.

[99] See Quantum technologies need a quantum energy initiative by Alexia Auffèves, November 2021 (10 pages), arXiv:2111.09241v2. This initiative is also a way for scientists to show they care early on about the environmental impact of technologies while they are designed and not as an aftereffect.

[100] See ORNL researchers advance performance benchmark for quantum computers, January 2020.

[101] See Scalable Benchmarks for Gate-Based Quantum Computers by Arjan Cornelissen et al, April 2021 (54 pages).

[102] See Application-Oriented Performance Benchmarks for Quantum Computing by Thomas Lubinski et al, October 2021 (33 pages).

[103] See P7131 - Standard for Quantum Computing Performance Metrics & Performance Benchmarking. It covers only gate-based quantum computing. See also Metrics & Benchmarks for Digital Quantum Computing by Robin Blume-Kohout (18 slides) and Summary of the IEEE Workshop on Benchmarking Quantum Computational Devices and Systems, 2019.

[104] See Benchmarking quantum co-processors in an application-centric, hardware-agnostic and scalable way by Simon Martiel, Thomas Ayral and Cyril Allouche, February 2021 (11 pages).

[105] See Science audiences, misinformation, and fake news by Dietram A. Scheufele and Nicole M. Krause, 2015 (8 pages).

[106] See this good example with the Don't fall for quantum hype short video by Sabine Hossenfelder. Also look at her excellent blog that is dedicated to debunking science hype.

[107] The difference between Technology Assessment and Responsible Research and Innovation methodologies is well laid out in Responsible innovation as a critique of technology assessment by Harro van Lente, Tsjalling Swierstra and Pierre-Benoît Joly, 2017 (9 pages).

[108] The challenges of responsible innovations are well described in Responsible Innovation. Managing the Responsible Emergence of Science and Innovation in Society, edited by R. Owen, J. Bessant, and M. Heintz, 2013 (293 pages), including What is 'Responsible' About Responsible Innovation? Understanding the Ethical Issues by Alexei Grinbaum, and Christopher Groves, 2013 (26 pages). There are many known examples of technology innovations which were widely deployed and had long-term negative outcomes: asbestos isolation in construction and lung cancers, DTT pesticides and their adverse environmental effects to wildlife and human health risks and chlorofluorocarbons refrigerants and their impact on the protective earth ozone layer.

[109] STEMs stands for science, technology, engineering and mathematics.

[110] See The 'second quantum revolution' is almost here. We need to make sure it benefits the many, not the few by Tara Robertson, June 2021.

[111] See Ethical Quantum Computing: A Roadmap by Elija Perrier, Centre for Quantum Software and Information, Sydney University of Technology, February 2021 (10 pages).

[112] See Reading the road: challenges and opportunities on the path to responsible innovation in quantum computing by Carolyn Ten Holter, Philip Inglesant and Marina Jirotka, University of Oxford, October 2021 (14 pages) and Asleep at the wheel? Responsible Innovation in quantum computing by Philip Inglesant, Carolyn Ten Holter, Marina Jirotka and Robin Williams, Universities of Oxford and of Edinburgh, 2021 (14 pages).

[113] See Responsible Innovation in Quantum Technologies applied to Defence and National Security by Philip Inglesant, Marina Jirotka and Mark Hartswood, University of Oxford, 2021 (13 pages).

[114] See Quantum Ethics | A Call to Action, February 2021 (13 mn video).

[115] See Quantum Computing Governance, World Economic Forum and Quantum Computing Governance Principles, Insight Report, January 2022 (35 pages). The two authors of this document, Kay Firth Butterfield and Arunima Sarkar have a strong background in machine learning and artificial intelligence.

[116] See Law & policy for the quantum age : a presentation by Chris Hoofnagle, February 2021 (58 mn). Berkeley Professor and the book Law and policy in the quantum age, by Chris Jay Hoofnagle and Simson L. Garfinkel, January 2020, free download (602 pages). Also Why Philosophers Should Care About Computational Complexity by Scott Aaronson, 2011 (53 pages). In What if Quantum Computing Is a Bust? by Chris Hoofnagle and Simon Garfinkel, the authors imagine what would happen if a quantum computing winter was to happen. They anticipate that the private sector will continue to invest in quantum computing. They view the private sector as a competitor to governments when it's actually their key technology providers.

[117] You can dig in many of the topics mentioned in this paper with the free to download ebook Understanding Quantum Technologies (September 2021, 836 pages, updated since then), that is also available on Arxiv and in paperback edition on Amazon sites (in two volumes).